\documentstyle[editedvolume,numreferences,epsf,psfig]{crckapb} 

\newcommand{\be}{\begin{equation}}
\newcommand{\ee}{\end{equation}}

\begin{opening}
\title{IMAGINATIVE COSMOLOGY}
 
\author{ROBERT H. BRANDENBERGER}
\institute{Physics Department, Brown University\\
           Providence, RI, 02912, USA}
 
\author{JO\~AO MAGUEIJO}
\institute{Theoretical Physics, The Blackett Laboratory, Imperial College\\
           Prince Consort Road, London SW7 2BZ, UK}

\end{opening}

\runningtitle{IMAGINATIVE COSMOLOGY}

\begin{document}

\begin{abstract}
We review{\footnote{Brown preprint BROWN-HET-1198, invited lectures at the International School on Cosmology, Kish Island, Iran, Jan. 22 - Feb. 4 1999, to be publ. in {\it Large Scale Structure Formation} (Kluwer, Dordrecht, 2000)}} a few off-the-beaten-track ideas in cosmology. They solve
a variety of fundamental problems;  also they are fun.
We start with a description of non-singular dilaton cosmology. In these
scenarios gravity is modified so that the Universe does not have a singular
birth. We then present a variety of ideas mixing string theory and cosmology.
These solve the cosmological problems usually
solved by inflation, and furthermore 
shed light upon the issue of the number
of dimensions of our Universe. We finally review several aspects of the
varying speed of light theory. We show how the horizon, flatness, and 
cosmological constant problems may be solved in this scenario. 
We finally present a
possible experimental test for a realization of this theory: a test in which
the Supernovae results are to be combined with recent evidence for redshift
dependence in the fine structure constant. 
\end{abstract}

\section{Introduction}

In spite of their unprecedented success at providing causal theories for the origin of structure, our current models of the very early Universe, in particular models of inflation and cosmic defect theories, leave several important issues unresolved and face crucial problems (see \cite{RB99} for a more detailed discussion). The purpose of this chapter is to present some imaginative and speculative ideas for early Universe cosmology which go beyond our present models and which address some of the puzzles unresolved by inflationary cosmology.

Let us recall four of the key conceptual problems of inflationary cosmology. The first is the fluctuation problem: without finely adjusting parameters in the particle physics model, the amplitude of the resulting cosmic microwave background (CMB) anisotropies is generically predicted to be far in excess of the measured values \cite{AFG}. What is missing is an in-built suppression mechanism for scalar metric fluctuations.

The second key problem of potential-driven inflation is the super-Planck scale physics problem: unless the period of inflation is very short, comoving length scales corresponding to today's large scale structure had a physical wavelength much smaller than the Planck length in the early stages of inflation. In the absence of an implementation of inflation in the context of a unified theory of Planck-scale physics, it is not justified to blindly extrapolate the predictions of simple scalar field toy models to the required small length scales.

Inflationary cosmology leaves the initial singularity problem of standard cosmology unresolved \cite{BV93}, and in this sense remains an incomplete theory of the
early Universe without a consistent initial value formalation. This is a further problem of inflation.

The Achilles heel of potential-driven inflationary cosmology is without doubt the cosmological constant problem. The mechanism of inflation relies on a transient cosmological constant generating the required exponential expansion of space, but can we trust this mechanism in the absence of an understanding of what sets the present value of the cosmological constant to (almost) zero?

In the following, we summarize some imaginative but very speculative ideas which address one or more of the above problems. At the present time, none of the ideas is well enough developed to be able to solve all of the problems and none forms a complete well-developed theory. Our hope, however, is that the ultimate theory of the early Universe will contain certain aspects of the mechanisms discussed below. As bi-products, these ideas point to the exciting possibility that the unified theory of all forces may explain the dimensionality of space \cite{BV89}, and yield an alternative to inflation for solving the flatness and horizon problems of standard cosmology without the need to invoke inflation \cite{AM98,Moffat}.

The outline of this chapter is as follows. In the next section, we show that adding a special class of higher derivative gravity terms to the usual low energy effective action for matter and gravity can lead to a nonsingular Universe. Applied to dilaton gravity, this construction may lead to a graceful exit in ``pre-big-bang" cosmology.

Although higher derivative gravity corrections to the Einstein action are a generic feature of any lower energy limit of a unified thoery, the specific terms used in the nonsingular Universe construction of Section 2 are rather ad hoc and not at the present time motivated by any fundamental microscopic theory. In Section 3 we study a few cosmological aspects of the one well-motivated candidate for a unified theory of all forces, string theory. In particular, we explore the role of target space duality in resolving cosmological singularities and in determining a preferential dimensionality of space.

Perhaps the craziest attempt to bypass inflation is the varying speed
of light cosmology \cite{AM98,Moffat}. In its current formulation the
theory allows for violations of Lorentz invariance. These are shown
to lead to a solution of the problems usually solved by inflation,
even if the matter content of the Universe satisfies the strong energy
condition. We outline some aspects of this theory in Section 4, and
in Section 5 we show how solutions to the cosmological problems
emerge in these theories. In Section 6 we then show how such a theory
may be confronted with experiment.

In a concluding section we digress on the future of imaginative 
cosmology (as opposed to mainstream, ``boring", cosmology).

\section{Nonsingular Dilaton Cosmology}

\subsection{Motivation}

The initial singularity is one of the outstanding problems of current
cosmological models. In standard big bang cosmology, the existence of
the initial singularity is an inevitable consequence of the
Penrose-Hawking theorems \cite{Penrose1965a,Hawking1967a}. While
potential-driven inflationary models such as chaotic
inflation \cite{Linde1983b} resolve many of the problems faced by
conventional cosmology, initial singularities are still generic, even
when stochastic effects are included \cite{BV93}.

It is generally hoped that string theory will lead to a resolution of the
singularity problem. A key feature of string theory is that the low energy degrees of freedom include besides the graviton the dilaton and an antisymmetric tensor field.  Gasperini and Veneziano
initiated a program known as ``pre-big-bang
cosmology" \cite{GasperiniET1992b} aimed at studying the cosmological consequences of the low energy effective action of string theory. To lowest order, and neglecting the antisymmetric tensor field, the action is given by
\be \label{E1}
S = -\frac{1}{2\kappa^2} \int{ d^4 x \sqrt{-g} \left\{
R - \frac{1}{2} (\nabla \phi)^2 + \cdots \right\}},
\ee
where $\phi$ is the dilaton, $\kappa^2 = 8\pi G = 8\pi m_{pl}^{-2}$, with
$G$ being the (4 dimensional) gravitational coupling, and $m_{pl}$
the Planck mass.  The field equations of pre-big-bang cosmology
exhibit a new symmetry, scale factor duality, which (in the Einstein
frame) maps an expanding Friedmann-Robertson-Walker (FRW) cosmology to
a dilaton-dominated contracting inflationary phase. This raises the
hope that it is possible to realize a nonsingular cosmology in which
the Universe starts out in a cold dilaton-dominated contracting phase,
goes through a bounce and then emerges as an expanding FRW
Universe (see Ref. \cite{Veneziano1998a} for a recent review
of pre-big-bang cosmology. Unfortunately, it has been shown that the two branches of pre-big-bang
cosmology cannot be smoothly connected within the above tree-level
action \cite{BruVen,EastherET1995a,KaloperET1995a,KaloperET1995b}. The
contracting dilaton-dominated branch has a future singularity, whereas
the expanding branch emerges from a past singularity.

A natural approach to resolving the singularity problem of general
relativity is to consider an effective theory of gravity which
contains higher order terms, in addition to the Ricci scalar
of the Einstein action. This approach is well motivated, since
we expect that any effective action for classical gravity obtained
from string theory, quantum gravity, or by integrating out matter
fields, will contain higher derivative terms. Thus, it is quite natural
to consider higher derivative effective gravity theories when studying
the properties of space-time at large curvatures.

Most higher derivative gravity theories have much worse singularity problems than Einstein's theory. However, it is not unreasonable to expect that in the fundamental theory of nature, be it string theory or some other theory, the curvature of space-time is limited. In Ref. \cite{Markov} the hypothesis was made that when the limiting curvature is reached, the geometry must approach that of a maximally symmetric space-time, namely de Sitter space. The question now becomes whether it is possible to find a class of higher derivative effective actions for gravity which have the property that at large curvatures the solutions approach de Sitter space. A {\it nonsingular Universe construction} which achieves this goal was proposed in Refs. \cite{MB92,BMS93}. It is based on adding to the Einstein action a particular combination of quadratic invariants of the Riemann tensor chosen such that the invariant vanishes only in de Sitter space-times. This invariant is coupled to the Einstein action via a Lagrange multiplier field in a way that the Lagrange multiplier constraint equation forces the invariant to zero at high curvatures. Thus, the metric becomes de Sitter and hence explicitly nonsingular.

\begin{figure}[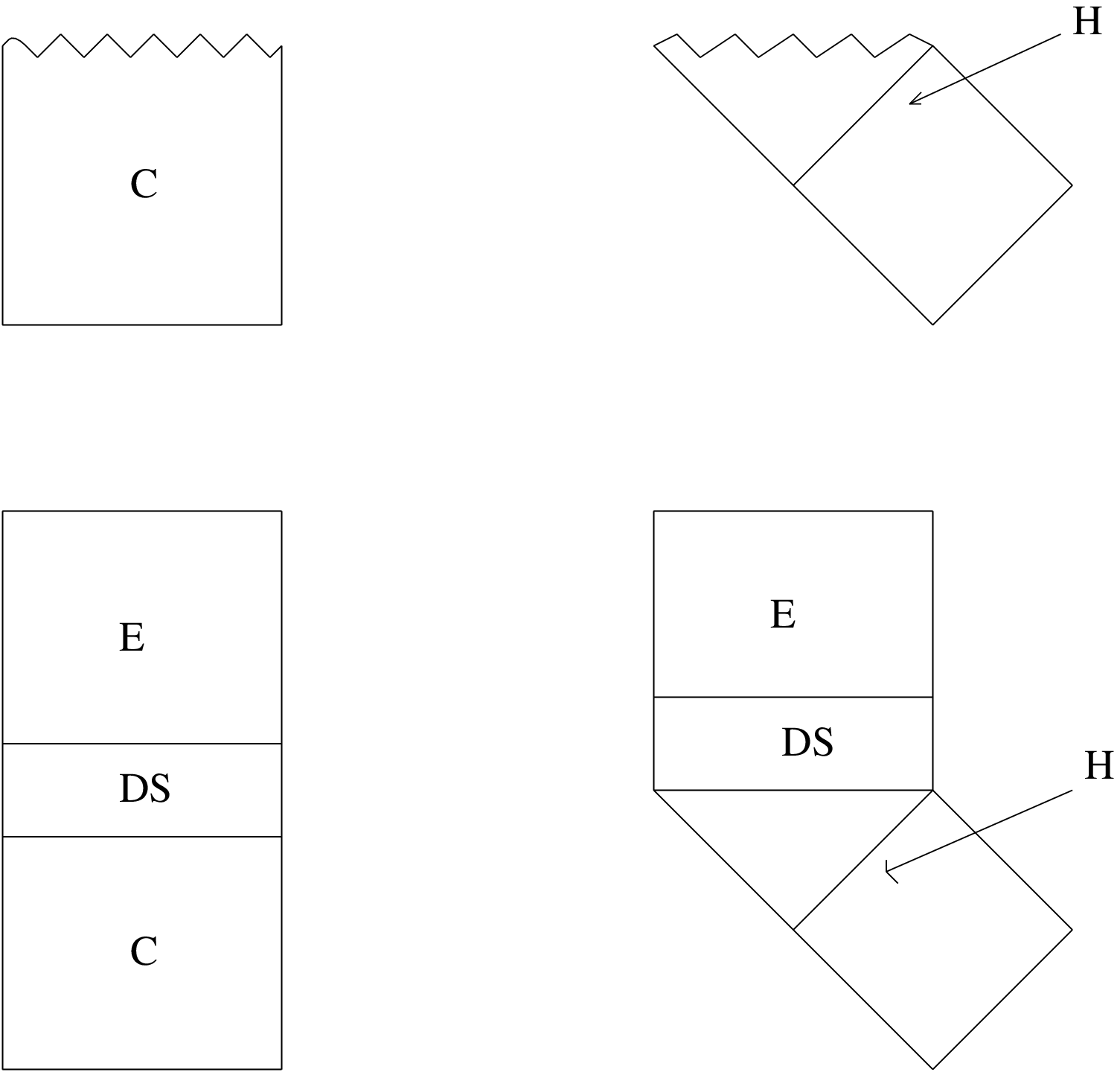]
\epsfysize=8cm \epsfbox{kishfig1.ps}   
\caption{Penrose diagrams for collapsing Universe (left) and black hole (right) in Einstein's theory (top) and in the nonsingular Universe (bottom). C, E, DS and H stand for contracting phase, expanding phase, de Sitter phase and horizon, respectively, and wavy lines indicate singularities.}
\end{figure}

If successful, the above construction will have some very appealing 
consequences.  Consider, for example, a collapsing spatially 
homogeneous Universe.  According to Einstein's theory, this Universe 
will collapse in a finite proper time to a final ``big crunch" singularity (top left Penrose diagram of Figure 1). 
In the new theory, however, the Universe will approach a de Sitter model as 
the curvature increases.  If the 
Universe is closed, there will be a de Sitter bounce followed by 
re-expansion (bottom left Penrose diagram in Figure 1).  Similarly, spherically 
symmetric vacuum solutions of the new equations of motion will presumably be nonsingular, i.e., black holes 
would have no singularities in their centers.  The structure of a 
large black hole would be unchanged compared to what is predicted by 
Einstein's theory (top right, Figure 1) outside and even slightly inside the horizon, since all curvature 
invariants are small in those regions.  However, for $r \rightarrow 0$ 
(where $r$ is the radial Schwarzschild coordinate), the solution 
changes and approaches a de Sitter solution (bottom right, Figure 1).  This 
would have interesting consequences for the black hole information 
loss problem (in two dimensions, this construction has been successfully realized \cite{TMB93}).

In this section, we review the {\it nonsingular Universe construction} of \cite{MB92,BMS93} and its applications to dilaton cosmology \cite{BEM98,EB99}.    

\subsection{An Analogy}

To motivate the {\it nonsingular Universe construction}, we turn to a well known example of a theory with a limited physical quantity, namely point particle motion in the theory of special relativity.  The transition from the Newtonian theory of point particle motion to 
the special relativistic theory transforms a theory with no bound on 
the velocity into one in which there is a limiting velocity, the speed 
of light $c$ (in the following we use units in which $\hbar = c = 1$).  
This transition can be obtained \cite{MB92} by starting with the Newtonian action of a point particle with world line $x(t)$:
\be
S_{\rm old} = \int dt {1\over 2} \dot x^2 \, , 
\ee
and adding a Lagrange multiplier $\varphi$ which couples to $\dot 
x^2$, the quantity to be made finite, and which has a potential 
$V(\varphi)$:
\be
S_{\rm new} = \int dt \left[ {1\over 2} \dot x^2 + \varphi \dot x^2 - 
V (\varphi) \right] \, .
\ee
{F}rom the Lagrange multiplier constraint equation
\be
\dot x^2 = {\partial V\over{\partial \varphi}} \, , 
\ee
it follows that $\dot x^2$ is limited provided $V(\varphi)$ increases 
no faster than linearly in $\varphi$ for large $|\varphi|$.  The small 
$\varphi$ asymptotics of $V(\varphi)$ is determined by demanding that 
at low velocities the correct Newtonian limit results:
\begin{eqnarray}
V (\varphi) \sim \varphi^2 & \> & {\rm as} \> |\varphi| 
\rightarrow 0 \, , \label{asympt1} \\
V (\varphi) \sim \varphi & \> & {\rm as} \> |\varphi| \rightarrow \infty 
\, . \label{asympt2}
\end{eqnarray}
Choosing the simple interpolating potential
\be
V (\varphi) = {2 \varphi^2\over{1 + 2 \varphi}} \, , 
\ee
the Lagrange multiplier can be integrated out, resulting in the well-known
action
\be
S_{\rm new} = {1\over 2} \int dt \sqrt{1 - \dot x^2}
\ee
for point particle motion in special relativity.

\subsection{Construction}

The procedure for obtaining a nonsingular Universe theory \cite{MB92} is based 
on generalizing the above Lagrange multiplier construction to gravity.  
Starting from the Einstein action, one can introduce a Lagrange 
multiplier $\varphi_1$ coupled to the Ricci scalar $R$ to obtain a 
theory with limited $R$:
\be \label{act1}
S = \int d^4 x \sqrt{-g} (R + \varphi_1 \, R + V_1 (\varphi_1) ) \, , 
\ee
where the potential $V_1 (\varphi_1)$ satisfies the asymptotic 
conditions (\ref{asympt1},\ref{asympt2}).

However, this action is insufficient to obtain a nonsingular gravity 
theory.  For example, singular solutions of the Einstein equations 
with $R=0$ are not affected at all.  The minimal requirements for a 
nonsingular theory are that {\it all} curvature invariants remain 
bounded and the space-time manifold is geodesically complete.  
Implementing the limiting curvature hypothesis \cite{Markov}, these conditions 
can be reduced to more manageable ones.  First, we choose one 
curvature invariant $I_1 (g_{\mu\nu})$ (e.g. $I_1 = R$ in (\ref{act1})) and demand that it be 
explicitely bounded by the construction of (\ref{act1}).  In a second step, we demand that as $I_1 (g_{\mu\nu})$ approaches its limiting value, the metric $g_{\mu\nu}$ approach 
the de Sitter metric $g^{DS}_{\mu\nu}$, a definite nonsingular metric 
with maximal symmetry.  In this case, all curvature invariants are 
automatically bounded (they approach their de Sitter values), and the 
space-time can be extended to be geodesically complete.

One can implement the second step of the above procedure by 
another Lagrange multiplier construction \cite{MB92}.  Consider a curvature 
invariant $I_2 (g_{\mu\nu})$ with the property that 
\be
I_2 (g_{\mu\nu}) = 0 \>\> \Leftrightarrow \>\> g_{\mu\nu} = 
g^{DS}_{\mu\nu} \, .
\ee
Next, introduce a second Lagrange multiplier field $\varphi_2$ which couples 
to $I_2$ and choose a potential $V_2 (\varphi_2)$ which forces $I_2$ 
to zero at large $|\varphi_2|$:
\be
S = \int d^4  x \sqrt{-g} [ R + \varphi_1 I_1 + V_1 (\varphi_1) + 
\varphi_2 I_2 + V_2 (\varphi_2) ] \, , 
\ee
with asymptotic conditions (\ref{asympt1},\ref{asympt2}) for $V_1 (\varphi_1)$ and conditions
\begin{eqnarray}
V_2 (\varphi_2) & \sim &{\rm const} \>\> {\rm as} \> | 
\varphi_2 | \rightarrow \infty \label{asympt3} \\
V_2 (\varphi_2) & \sim & \varphi^2_2 \>\> {\rm as} \> |\varphi_2 | 
\rightarrow 0 \, , \label{asympt4}
\end{eqnarray}
for $V_2 (\varphi_2)$.  The first constraint forces $I_2$ to zero, the 
second is required in order to obtain the correct low curvature limit.

These general conditions are reasonable, but not sufficient in order 
to obtain a nonsingular theory.  It must still be shown that all 
solutions are well behaved, i.e., that they asymptotically reach the 
regions $|\varphi_2| \rightarrow \infty$ of phase space (or that 
they can be controlled in some other way).  This must be done for a 
specific realization of the above general construction.

At the moment we are only able to find an invariant $I_2$ which 
singles out de Sitter space by demanding $I_2 = 0$ provided we assume 
that the metric has special symmetries.  The choice
\be \label{inv}
I_2 = (4  R_{\mu\nu} R^{\mu\nu} - R^2 + C^2)^{1/2} \, ,
\ee
singles out the de Sitter metric among all homogeneous and isotropic 
metrics (in which case adding $C^2$, the Weyl tensor square, is 
superfluous), all homogeneous and anisotropic metrics, and all 
radially symmetric metrics.

As a specific example one can consider the action \cite{MB92,BMS93}
\be \label{act2}
S = \int d^4 x \sqrt{-g} \left[ R + \varphi_1 R - (\varphi_2 + 
{3\over{\sqrt{2}}} \varphi_1) I_2^{1/2} + V_1 (\varphi_1) + V_2 
(\varphi_2) \right] 
\ee
with
\begin{eqnarray}
V_1 (\varphi_1) & = & 12 \, H^2_0 {\varphi^2_1\over{1 + \varphi_1}} \left( 1 
- {\ln (1 + \varphi_1)\over{1 + \varphi_1}} \right) \\
V_2 (\varphi_2) & =  & - 2 \sqrt{3} \, H^2_0 \, {\varphi^2_2\over{1 + 
\varphi^2_2}} \, .
\end{eqnarray}

When restricted to the set of homogeneous and isotropic space-times, the equations of motion which follow from the action (\ref{act2}) are not hard to
analyze \cite{MB92,BMS93}. They consist of the two Lagrange multiplier constraint equations and the $g_{00}$ equation. The Lagrange multiplier constraint equations can be used to eliminate the Hubble parameter and its derivative in favor of functions of the Lagrange multiplier fields. The remaining one dynamical equation becomes an equation for the flow in the
$(\varphi_1, \varphi_2)$ phase space. It can be shown that all solutions are either periodic about Minkowski space-time $(\varphi_1, \varphi_2) = (0, 0)$ or else asymptotically approach de Sitter space ($|\varphi_2 | \rightarrow \infty$).

One of the most interesting properties of this theory is asymptotic 
freedom \cite{BMS93}, i.e., the coupling between matter and gravity goes to 
zero at high curvatures.  It is easy to add matter (e.g., dust, 
radiation or a scalar field) to the gravitational action in the standard way.
One finds that in the asymptotic de Sitter regions, the trajectories of 
the solutions projected onto the $(\varphi_1, \, \varphi_2)$ plane are 
unchanged by adding matter.  This applies, for example, in a phase of de Sitter 
contraction when the matter energy density is increasing exponentially 
but does not affect the metric.  The physical reason for asymptotic 
freedom is obvious: in the asymptotic regions of phase space, the 
space-time curvature approaches its maximal value and thus cannot be 
changed even by adding an arbitrarily high matter energy density.

In retrospect, it is not necessary to explicitly bound an invariant $I_1$ to obtain a nonsingular Universe. In fact, $\varphi_1$ can be eliminated from the action (\ref{act2}) without affecting the nonsingularity of the resulting cosmological solutions, as was explicitly demonstrated in \cite{BMS93}. Hence, in our application to dilaton cosmology we will only keep a single Lagrange multiplier field $\varphi_2$ which we will henceforth denote by $\psi$.

\subsection{Application to Dilaton Cosmology}

The starting point is the action (\ref{E1}) for dilaton gravity
(written in the Einstein frame) to which we \cite{BEM98} add the higher derivative
term given by $I_2$, in analogy to what was done above in the absence of the
dilaton. To be specific, we choose minimal coupling of $I_2$ to the dilaton.
\be  \label{E4}
S = \frac{-1}{2\kappa^2} \int{ d^4 x \sqrt{-g} \left\{
R - \frac{1}{2} (\nabla \phi)^2 + {1 \over {\sqrt{12}}} \psi  I_2 + V(\psi) 
\right\}}.
\ee
 
Restricted to a homogeneous and isotropic metric of the form
\be
ds^2 = dt^2 - a(t)^2 \bigl( {1 \over {1 - kr^2}}dr^2 + r^2 d\Omega^2
\bigr) \, ,
\ee
where $d\Omega^2$ is the metric on $S^2$, the equations of motion
resulting from (\ref{E4}) become
\begin{eqnarray}
\dot{\psi} \, &=& \, - 3 H \psi \, + 6 H \, - \, {1 \over H}
\bigl( {1 \over 2} \chi^2 \, + \, V \bigr), \nonumber \\
\dot{H} \, &=& \, \ -V(\psi)^{\prime}, \nonumber\\
\dot{\chi} \, &=& \, - 3 H \chi,  \label{E7}
\end{eqnarray} 
where dots denote derivatives with respect to time, $t$, $\chi = \dot{\phi}$ and a prime ($\prime$) signifies the derivative
with respect to $\psi$. Since the initial goal is to construct a spatially flat, bouncing Universe, we have set the curvature constant $k = 0$. 
 
Consider next the criteria which the potential
$V(\psi)$ must satisfy. At small curvatures, the terms in the action
(\ref{E4}) which depend on $\psi$ must be negligible compared to the
usual terms of dilaton gravity. This is ensured by
demanding
\be  
V(\psi) \, \sim \, \psi^2 \,\,\,\,\,\, |\psi| \rightarrow 0 \label{E8}
\ee
as the region of small $|\psi|$ will correspond to the low curvature
domain \cite{BMS93}. In order to implement the
limiting curvature hypothesis, the invariant $I_2$ must tend to zero,
and thus the metric $g_{\mu \nu}$ will tend to a de Sitter metric at
large curvatures, i.e. for $|\psi| \rightarrow \infty$.  From the
variational equation with respect to $\psi$, it is obvious that this
requires
\be
V(\psi) \, \rightarrow \, {\rm const} \,\,\,\,\,\, |\psi| \rightarrow
\infty \, . \label{E9}
\ee

Conditions (\ref{E8}) and (\ref{E9}) are the same as those  (\ref{asympt3},\ref{asympt4}) used in the previous subsection, but they do not
fully constrain the potential.  In order to obtain a bouncing solution
in the absence of spatial curvature it is necessary to add a third
criterion: the equations must allow a configuration with $H = 0$ and
$\psi \neq 0$. From the equation of motion for $\psi$ in (\ref{E7}) it
follows that $V(\psi)$ must become negative, assuming that it is
positive for small $|\psi|$. Let $\psi_b$ denote the nontrivial zero
of $V(\psi)$: 
\be V(\psi_b) \, = \, 0 \, . \label{E10}
\ee 
In the absence of the dilaton, $\psi_b$ will correspond to the
value of $\psi$ at the bounce. In the presence of $\phi$, the value of
$|\psi|$ at the bounce will depend on $\chi$ and will be larger than
$|\psi_b|$.
A simple potential which satisfies the conditions (\ref{E8}), (\ref{E9})
and (\ref{E10}) is \cite{BEM98}
\be
V(\psi) = \frac{\psi^2 - \frac{1}{16}\psi^4}{1+\frac{1}{32} \psi^4}.
\ee
 
The conditions, (\ref{E8}) - (\ref{E10}), on the potential $V(\psi)$
discussed in the previous section are necessary but not sufficient to
obtain a nonsingular cosmology. These conditions ensure that all
solutions which approach large values of $|\psi|$ are nonsingular, but
the possibility of geodesic incompleteness for solutions which always
remain within the small $|\psi|$ region remains to be studied. If $\chi = 0$ then the trajectories of the solutions in the phase plane $(\psi, H)$ can be studied quite easily analytically \cite{BEM98} and one can thereby explicitly demonstrate the absence of singularities. As a new feature (and as required in pre-big-bang cosmology), the theory admits spatially flat bouncing
solutions.

There are several special points and curves on the phase plane $(\psi, H)$.
First, the point $(\psi, H) = (0, 0)$ corresponds to Minkowski space-time.
The potential $V(\psi)$ vanishes at this point, but it also vanishes at
the points
\be
\psi_b = \pm 4.
\ee
Trajectories which pass through the phase plane points $(\psi_b, 0)$ correspond
to spatially flat bouncing cosmologies.
To demonstrate that the point $(\psi, H) = (4, 0)$ is in fact a bounce,
one expands the $\psi$ equation of motion near $H = 0$, which yields
\be
H {\dot \psi} \, = \, - V \, . \label{E14}
\ee
When crossing the $H = 0$ axis, the sign of ${\dot \psi}$ changes.
Contracting solutions with $2 < \psi < 4$ have ${\dot \psi} > 0$ and
approach the point $(4, 0)$ in finite time since ${\dot H}$ is positive
and does not tend to zero, and emerge for $H > 0$ as expanding trajectories
with decreasing curvature (since ${\dot \psi} < 0$). The trajectories in the phase plane are symmetric about the
$H = 0$ axis, except that the time arrows are reversed.

By expanding the equations near the origin of the phase plane one can demonstrate the existence of critical lines, lines with 
\be
{{d \psi} \over {dH}} = 0
\ee
located at
\be
\psi_c(H) \, = \, \pm \, \sqrt{6} H \, 
\ee
(for large values of $\vert H \vert$ the critical lines approach the lines $\psi = 2$). Focusing on the contracting solutions, trajectories which lie above the 
critical line have
${\dot H} < 0$ and ${\dot \psi} > 0$. These
trajectories thus are directed towards the line $\psi = 2$ where ${\dot H}$
changes sign. Provided they do not cross the critical line, solutions
which start out in this region of the phase plane are thus candidates
for spatially flat bouncing Universes.
Solutions below the critical line have ${\dot \psi} < 0$ and will
hence not exhibit a bounce. 

There is a separatrix line between phase plane trajectories which
start near the origin and which reach $\psi = 2$ (and are thus
candidates for a bouncing Universe) and those trajectories which cross
the critical line $\psi_c(H)$ and turn around, i.e. become solutions
with ${\dot \psi} < 0$. The precise location of this separatrix is
relevant to the initial condition problem of pre-big-bang cosmology (see e.g. \cite{KLB}). The analysis of \cite{BEM98} yields the estimate 
\be \label{E22}
|H| \, = \, {1 \over c} \psi^2 \,\,\, {\rm with} \, c \geq 4
\sqrt{2}
\, .
\ee
with $c \gg 4$ for the location of the separatrix near the origin of phase space.  

The only region of phase space where singular solutions might occur are the critical lines mentioned above. However, one can show \cite{BEM98} that these critical lines not attractors, and hence trajectories
peel away from these lines as $\vert H \vert \rightarrow \infty$ and tend to the asymptotic de Sitter region.

\begin{figure}[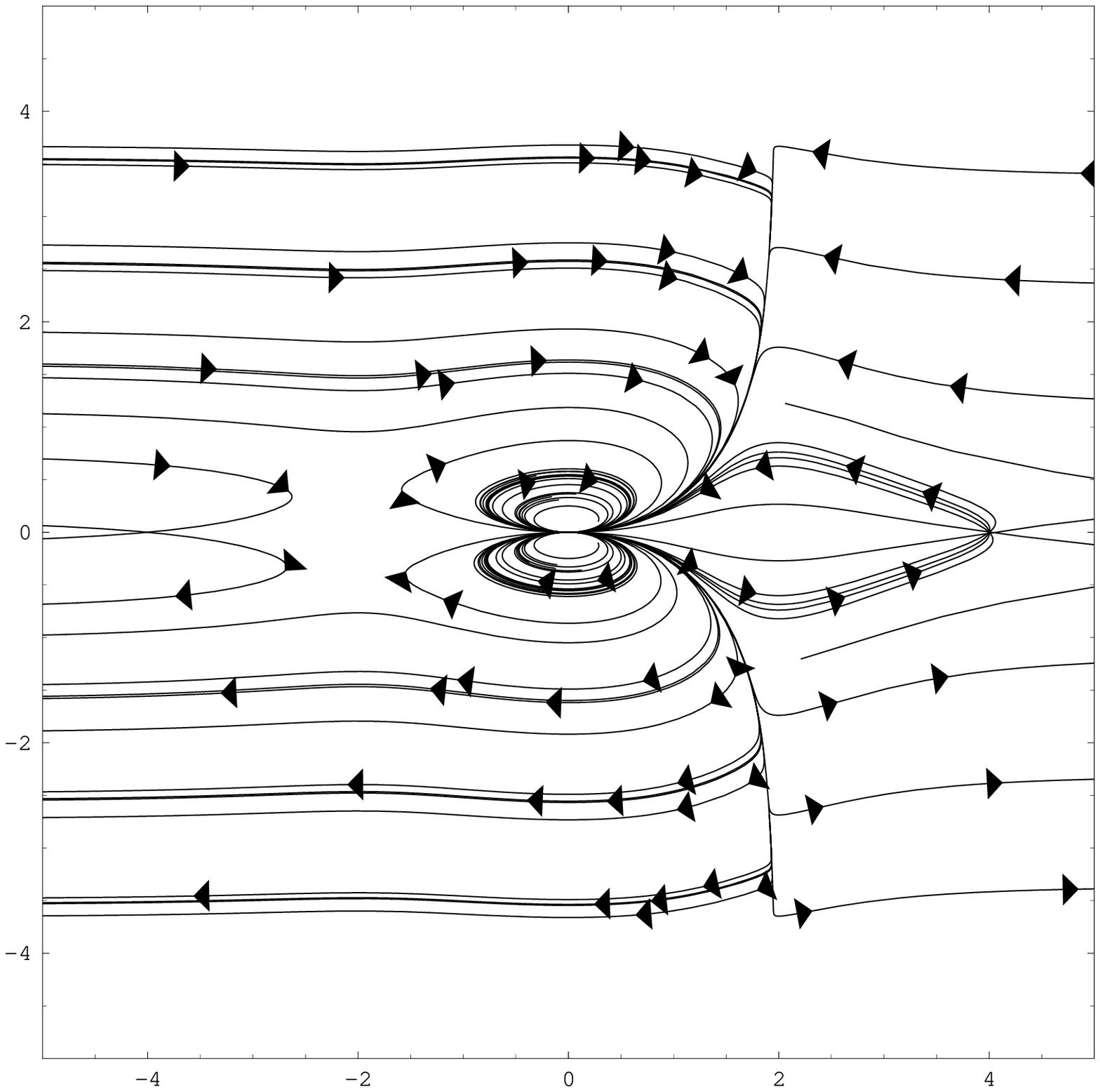]
\epsfysize=8cm \epsfbox{kishfig2.ps}
\caption{The phase portrait for solutions of the equations of
motion, with the dilaton kinetic energy ($\chi$) set to zero is
shown. $H$ is plotted on the vertical axis, while $\psi$
is plotted on the horizontal.}
\end{figure}

The equations of motion (\ref{E7}) were solved
numerically. The resulting phase diagram is shown in Figure 2. The
absence of singular solutions is manifest. In the asymptotic regions
$|\psi| \gg 4$, all solutions tend to de Sitter space. The critical
lines are seen to repel contracting solutions towards the asymptotic
de Sitter regions. The most interesting class of solutions are those
corresponding to a cosmological bounce. As predicted, they form a set
of finite measure among solutions which start out close to the origin
of the phase plane. There are also solutions which ``oscillate'' about
Minkowski space-time.

In the presence of the dilaton, the phase space becomes three
dimensional ($\psi(t) , \,  H(t) , \, \chi(t)$) and therefore more
difficult to discuss analytically. Note that in the Einstein
frame the dilaton corresponds to a homogeneous free massless scalar
field.
Thus, it is
already clear from the analysis of \cite{BMS93} that
the dilaton will not introduce any singularities into the system. In
fact, from the equation of motion for $H$ (see (\ref{E7})) it follows
that, in the region of large $|H|$, the presence of $\chi$ will not change the phase space
trajectories projected onto the ($\psi, H$) plane, a manifestation of the property of {\it asymptotic freedom} mentioned above. However, the
presence of $\chi$ will greatly accelerate the time evolution of
$\psi$ on the given ($\psi, H$) trajectory. This is easy to see for
the contracting de Sitter solutions, since in this case the $\chi$
equation of motion (see (\ref{E7})) leads to exponential growth of
$\chi$ which, inserted into the $\psi$ equation of motion,
demonstrates that at large values of $|\psi|$, the $\chi^2$ term
dominates the evolution of $\psi$.

Consider now the effect of $\chi$ on the trajectories in the ($\psi,
H$) phase plane. The role of the lines $\psi = \pm 2$ remains
unchanged: they correspond to maxima of $|H|$ for any given
trajectory. However, the condition for the bounce changes. Instead of
$V(\psi_b) = 0$, it now follows from (\ref{E7}) that the condition
becomes
\be \label{E31}
{1 \over 2} \chi^2 \, + \, V(\psi_b) \, = \, 0 \, .
\ee
Hence, $|\psi_b|$ is shifted to a larger (and dilaton-dependent)
value. Note that for very large large values of $|\chi|$ for which the
above equation has no solutions there will be no bounce. 

The presence of $\chi$ also changes the critical lines. For a fixed
value of $\psi$, the condition ${\dot \psi} = 0$ which determines the
critical line occurs at a larger value of $|H|$ than it does in the
absence of $\chi$, as can easily be seen from the $\psi$ equation of
motion (see (\ref{E7}).  In addition, for fixed initial ($\psi, H$)
the value of ${\dot \psi}$ for contracting solutions is larger with
$\chi \neq 0$ than with $\chi = 0$. It thus follows that adding a small value
of $\chi$ increases the range of initial conditions in the ($\psi, H$)
plane which lead to a bounce. 

It can be argued that collapsing solutions in the presence of a
small $|\chi|$ quite generically lead to a spatially flat bouncing
Universe. Consider initial conditions with small but
positive $\chi$ which in the ($\psi, H$) plane lie below the
separatrix line and which initially ``oscillate'' about Minkowski
space-time. These trajectories do not bounce
upon reaching $H = 0$ but start another cycle with $H < 0$. Since $H
\leq 0$ at all times, $\chi(t)$ is increasing. Eventually, $\chi(t)$
reaches a sufficiently large value such that the trajectory crosses
the ``separatrix'' in the ($\psi, H$) plane and evolves past $\psi =
2$ to a successful bounce.  An example of such a trajectory is shown
in Figure 3. The top left panel shows the evolution of $H$ as a function of
time, the top right depicts that of $a(t)$, the bottom left that of $\chi(t)$ and the bottom right that of $\psi(t)$.

\begin{figure}[htbp]
\begin{center}
\begin{tabular}{ll}
\epsfysize=5cm \epsfbox{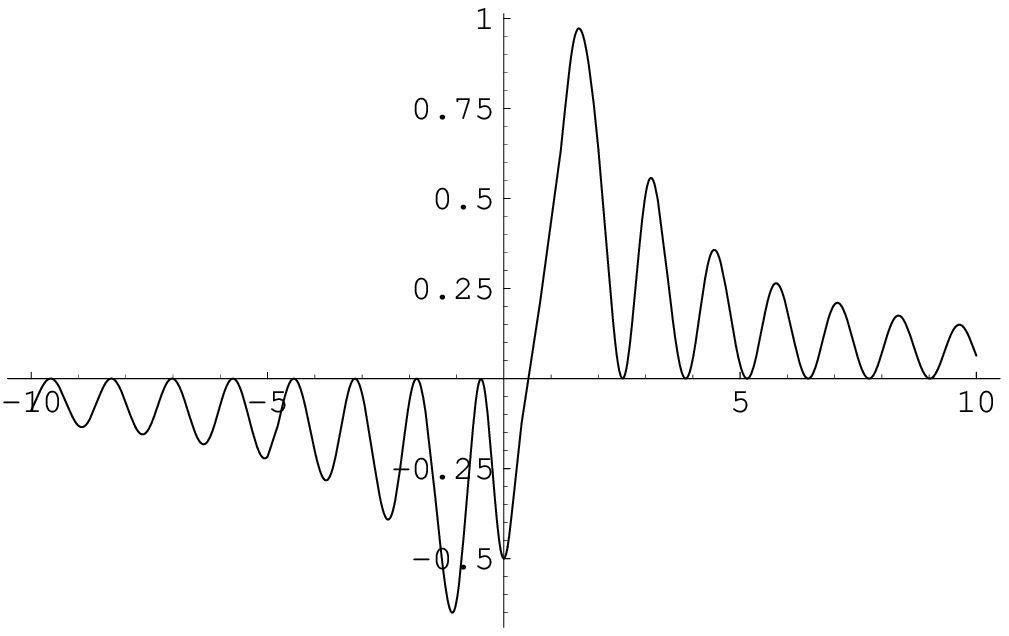} &
\epsfysize=5cm \epsfbox{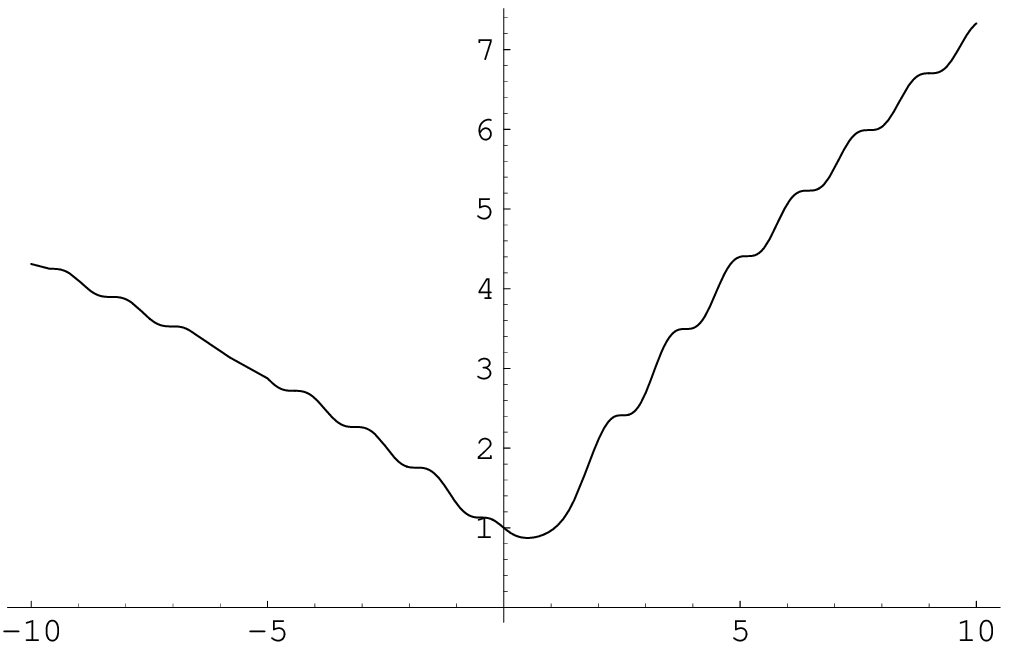} \\
\epsfysize=5cm \epsfbox{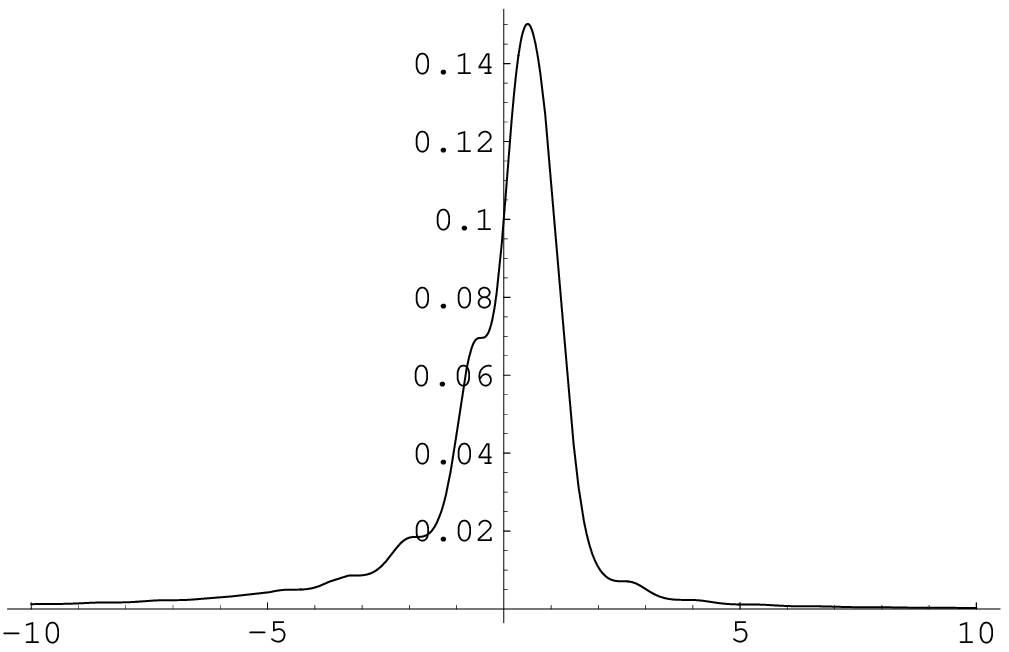} &
\epsfysize=5cm \epsfbox{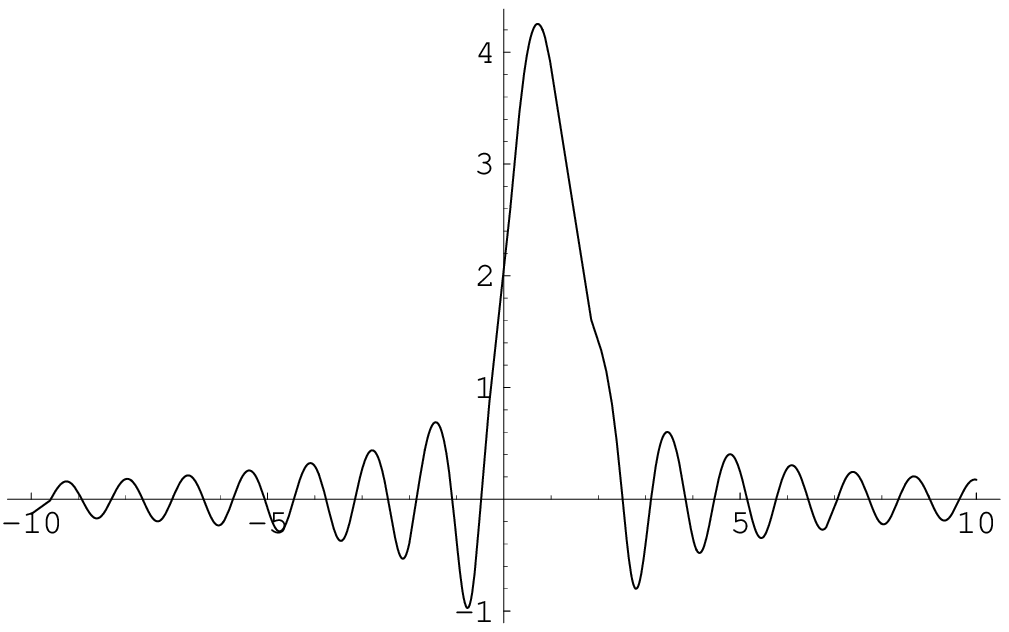} 
\end{tabular}
\end{center}
\caption{A specific bouncing solution with a non-trivial
contribution from the dilaton is plotted. The left top panel shows the evolution
of $H$ as a function of time, the right top panel depicts that of $a(t)$, on the bottom left the evolution of $\chi(t)$ is shown and on the bottom right that of $\psi(t)$.}
\end{figure}

It is now obviously possible to construct initial conditions in which
the Universe is initially dominated by the dilaton and contracting
towards a bounce. For example, take the initial conditions for
($\psi, H, \chi$) to be those of the solution in Figure 3 at the end
of the cycle preceding the bounce. Thus, the nonsingular Universe construction
provides a successful implementation of the evolution postulated in
pre-big-bang cosmology. Note that after the bounce, the dilaton
rapidly becomes irrelevant to the evolution in the expanding phase.
Note, in particular, that the dilaton tends to a constant. This feature
is different from what happens in other attempts to achieve a branch change
where the dilaton continues to grow in the expanding phase.

\subsection{Discussion}

It has been shown that a class of higher derivative extensions of 
Einstein gravity exists \cite{MB92,BMS93} for which all homogeneous and isotropic cosmological
solutions are nonsingular.  This class of models is very special.  Most higher 
derivative theories of gravity have, in fact, much worse singularity 
properties than the Einstein theory.  What is special about our class 
of theories is that they are obtained using a well motivated Lagrange 
multiplier construction which implements the limiting curvature 
hypothesis.   

The construction of \cite{MB92,BMS93} immediately carries over to dilaton gravity \cite{BEM98} and the resulting action solves the graceful exit problem of pre-big-bang cosmology, giving automatically a spatially flat bounce interpolating between the contracting dilaton-dominated epoch and the expanding FRW phase. From the point of view of string theory and pre-big-bang cosmology it is more natural to work in the string frame, rather than in the Einstein frame (as was done in \cite{BEM98}). However, the nonsingular Universe construction can also be applied to dilaton gravity in the string frame, with identical results \cite{EB99}.

It is to be expected that the nonsingular Universe construction with the invariant $I_2$ given by (\ref{inv}) also will regulate the singularities of black holes and of anisotropic homogeneous cosmologies. However, with the exception of the case of the two-dimensional black hole \cite{TMB93} the analysis has not been completed.

By construction, all solutions are de Sitter at high curvature.  Thus, 
the theories automatically have a period of inflation (driven by the 
gravity sector in analogy to Starobinsky inflation \cite{AS80}) in the 
early Universe.
A very important property of these theories is asymptotic freedom.  This 
means that the coupling between matter and gravity goes to zero at 
high curvature, and might lead to an automatic suppression mechanism 
for scalar fluctuations. Thus, the nonsingular Universe construction has the potential of solving two of the problems of potential-driven inflationary cosmology listed in Section 1.

\section{Elements of String Cosmology}

\subsection{Motivation}

In the previous section we studied effective actions for the space-time 
metric and the dilaton fields which might arise in the intermediate energy regime of string theory.  However, it is also of interest to 
explore the predictions of string theory which depend specifically on 
the ``stringy" aspects of the theory and which are lost in any field 
theory limit.  It is to a description of a few of the string-specific 
cosmological aspects to which we turn in this section.

\subsection{Implications of Target Space Duality}

Target space duality \cite{tdual} is a symmetry specific to string theory.  
As a simple example, consider a superstring background in which all 
spatial dimensions are toroidally compactified with equal radii.  Let 
$R$ denote the radius of the torus.

The spectrum of string states is spanned by oscillatory modes which 
have energies independent of $R$, by momentum modes whose energies 
$E_n$ (with integer $n$) are
\be
E_n = {n\over R} \, , 
\ee
and by winding modes with energies $E^\prime_m$ ($m$ integer)
\be
E^\prime_m = mR \, . 
\ee

Target space duality is a symmetry between two superstring theories, 
one on a background with radius $R$, the other on a background of 
radius $1/R$, under which winding and momentum modes are interchanged.

Target space duality has interesting consequences for string 
cosmology \cite{BV89}.  Consider a background with adiabatically changing 
$R(t)$.  While $R(t) \gg 1$, most of the energy in thermal equilibrium 
resides in the momentum modes.  The position eigenstates $|x >$ are 
defined as in quantum field theory in terms of the Fourier transform 
of the momentum eigenstates $|p >$
\be
|x > = \sum\limits_p e^{i x \cdot p} |p > \, . 
\ee
However, for $R (t) \ll 1$, most of the energy flows into winding 
modes, and it takes much less energy to measure the ``dual distance" 
$| \tilde x >$ than $|x >$, where 
\be
| \tilde x > = \sum\limits_w e^{i \tilde x \cdot w} | w > 
\ee
is defined in terms of the winding modes $| w>$.

We conclude that target space duality in string theory leads to a 
minimum physical length in string cosmology.  As $R(t)$ decreases 
below 1, the measured length starts to increase again.  This could 
lead to a bouncing or oscillating cosmology \cite{BV89}.

It is  well known that for strings in thermal equilibrium there is a 
maximal temperature, the Hagedorn temperature \cite{Hagedorn}.  Target space 
duality implies that in thermal equilibrium the temperature in an 
adiabatically varying string background begins to decrease once $R(t)$ 
falls below 1:
\be
T \left({1\over R} \right) = T(R) \, . 
\ee
Thus, the $T(R)$ curve in string cosmology is nonsingular and very 
different from its behavior in standard cosmology.  For further 
discussions of the thermodynamics of strings see, e.g., \cite{MT87} and 
\cite{DST89} and references therein.

\subsection{Strings and Space-Time Dimensionality}

Computations \cite{BV89} using the microcanonical ensemble show that for 
all spatial directions compactified at large total energy $E$, the 
entropy $S$ is proportional to $E$:
\be
S = \beta_H E \, , 
\ee
with $\beta_H$ denoting the inverse of the Hagedorn temperature $T_H$.  
Thus, the $E(R)$ curve in string cosmology is very different from the  
corresponding curve in standard cosmology.

For large $R \gg 1$, most of the energy in a gas of strings in thermal 
equilibrium will flow into momentum modes, and the thermodynamics will 
approach that of an ideal gas of radiation for which
\be
E (R) \sim {1\over R} \, . 
\ee
By duality, for small $R$
\be \label{winden}
E (R) \sim R \, . 
\ee

If, however, for some reason the string gas falls out of equilibrium, 
the $E(R)$ curve will look very different.  Starting at $R= 1$ with a 
temperature approximately equal to $T_H$, a large fraction of the 
energy will reside in winding modes.  If these winding modes cannot 
annihilate, thermal equilibrium will be lost, and the energy in 
winding modes will increase linearly in $R$, and thus for large $R$: 
\be
E (R) \sim R \, . 
\ee

Newtonian intuition indicates that out of equilibrium winding modes 
with an energy relation (\ref{winden}) will prevent the background space from 
expanding \cite{BV89}.  The equation of state corresponding to a gas of 
straight strings is
\be
p = - {1\over N} \rho 
\ee
where $p$ and $\rho$ denote pressure and energy density, respectively, and
$N$ is the number of spatial dimensions.
According to standard general relativity, an equation of state with 
negative pressure will lead to more rapid expansion of the background.  
It turns out that the Newtonian intuition is the correct one and that 
general relativity gives the wrong answer \cite{V91,TV92}.  At high densities, 
the specific stringy effects -- in particular target space duality 
 -- become crucial.

The Einstein action violates duality.  In order to restore duality, it 
is necessary to include the dilaton in the effective action for the 
string background.  Recall that the action for dilaton gravity can be written as
\be \label{dilact}
S = \int d^{N+1} x \sqrt{-g} e^{-2 \phi} [ R+ 4 (\nabla \phi)^2 ] 
\ee
where $\phi$ is the dilaton (rescaled compared to (\ref{act1})).  It is convenient to use new fields $\varphi$ and $\lambda$ 
defined by
\be
a (t) = e^{\lambda t} 
\ee
and
\be
\varphi = 2 \phi - N \lambda \, . 
\ee
The action (\ref{dilact}) has the duality symmetry
\be
\lambda \rightarrow - \lambda, \> \varphi \rightarrow \varphi \, . 
\ee

The variational equations of motion derived from (\ref{dilact}) for a 
homogeneous and isotropic model are \cite{V91,TV92}
\begin{eqnarray}
\dot \varphi^2 &=& e^\varphi E + N \dot \lambda^2 \\
\ddot \lambda - \dot \varphi \dot \lambda &=& {1\over 2} e^\varphi P \\
\ddot \varphi &=& {1\over 2} e^\varphi E + N \dot \lambda^2 \, , 
\end{eqnarray}
where $P$ and $E$ are total pressure and energy, respectively.  For a 
winding mode-dominated equation of state (and neglecting friction 
terms) the equation of motion for $\lambda (t)$ becomes
\be
\ddot \lambda = - {1\over{2N}} e^\varphi E(\lambda) \, , 
\ee
which corresponds to motion in a confining potential.  Hence, winding modes prevent the background toroidal 
dimensions from expanding.

These considerations may be used to put forward the conjecture \cite{BV89} 
that string cosmology will single out three as the maximum number of 
spatial dimensions which can be large ($R \gg 1$ in Planck units).  
The argument proceeds as follows.  Space can, starting from an initial 
state with $R \sim 1$ in all directions, only expand if thermal 
equilibrium is maintained, which in turn is only possible if the 
winding modes can annihilate.  This can only happen in at most three spatial
 dimensions (in a higher number the probability for 
intersection of the world sheets of two strings is zero).  In the 
critical dimension for strings, $N=3$, the evolution of a string gas 
has been studied extensively in the context of the cosmic string 
theory (see e.g., Refs. \cite{VS94}, \cite{HK95} and \cite{RB94} for recent reviews).  The winding 
modes do, indeed, annihilate, leaving behind a string network with 
about one winding mode passing through each Hubble volume.  Thus, in 
string cosmology only three spatial dimensions will become large 
whereas the others will be confined to Planck size by winding modes.
Crucial aspects of this scenario have been confirmed in numerical simulations with cosmic strings in three and four spatial dimensions \cite{MS96}.


\section{A varying speed of light}
One of the most controversial alternatives to inflation is the
varying speed of light (VSL) theory. This idea was initially proposed by
Moffat \cite{Moffat}, and has recently attracted a lot of attention
\cite{AM98,barrow,vsl1,vsl2,vsl3,cly1,cly2,ot,drummond}. In the 
same way inflation is not a model, but a vast class of models, 
the varying speed of light theory may also receive very different 
realizations. Here we present some possible VSL models. 

\subsection{The inspiration}
\begin{figure}
\centerline{\psfig{file=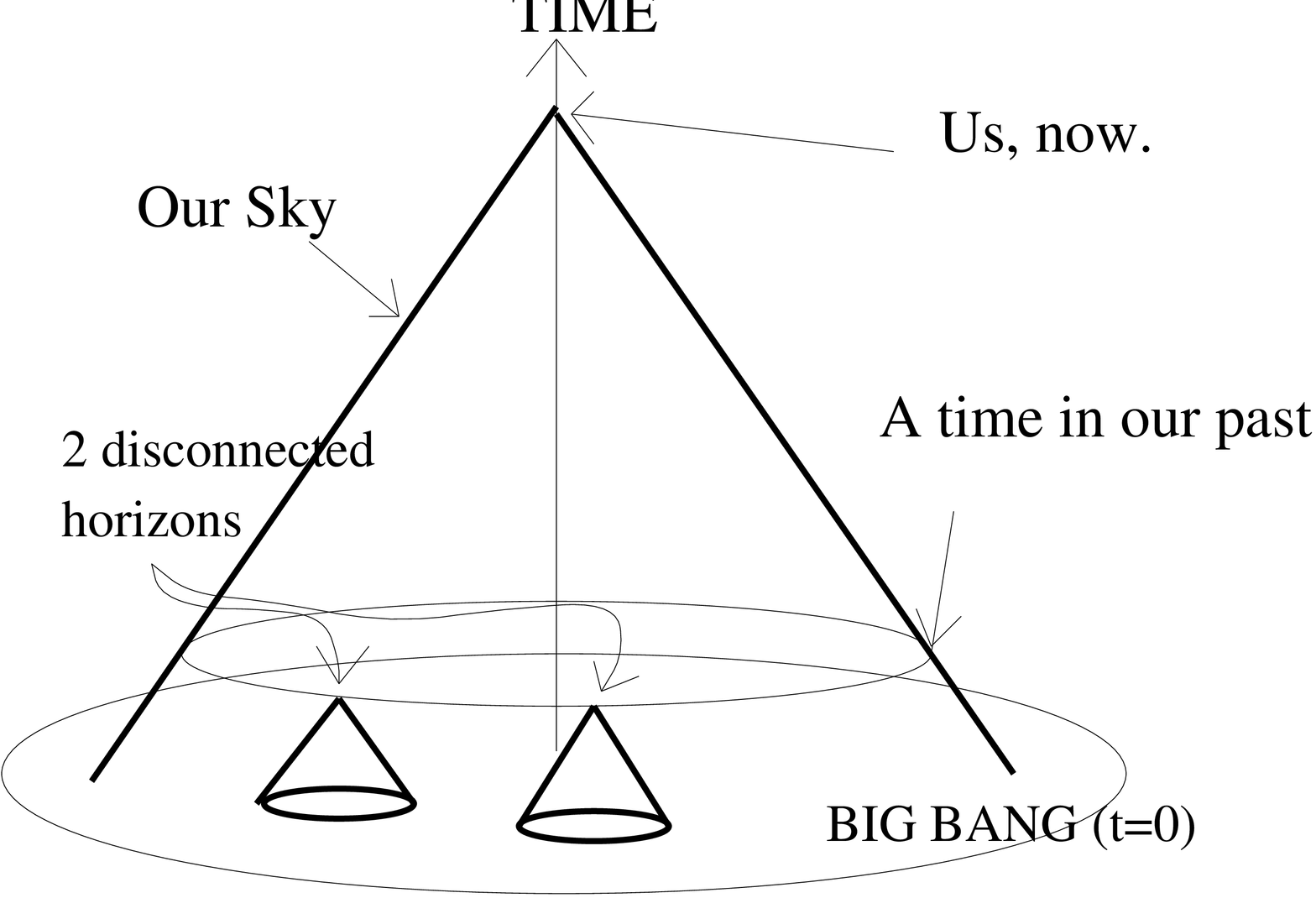,width=6 cm,angle=-90}}
\caption{Conformal diagram (light at $45^\circ$) showing the
horizon structure in the SBB model. Our past light cone contains 
regions outside each others' horizon.}
\label{fig1}
\end{figure}

The insight leading to VSL concerns the horizon problem.
At any given time any observer
can only see a finite region of the Universe, with comoving radius
$r_h=c\eta$, where $\eta$ denotes conformal time, and 
$c$ the speed of light. Since the horizon
size increases with time we can now observe many regions in our past 
light cone which are causally disconnected, that is, outside each others'
horizon (see Fig.~\ref{fig1}). 
The fact that these regions have the same properties (eg.
Cosmic Microwave background temperatures equal 
to a few parts in $10^5$) is puzzling
as they have not been in physical contact. This is a mystery one may
simply relegate to the setting up of initial conditions in our Universe.

\begin{figure}
\centerline{\psfig{file=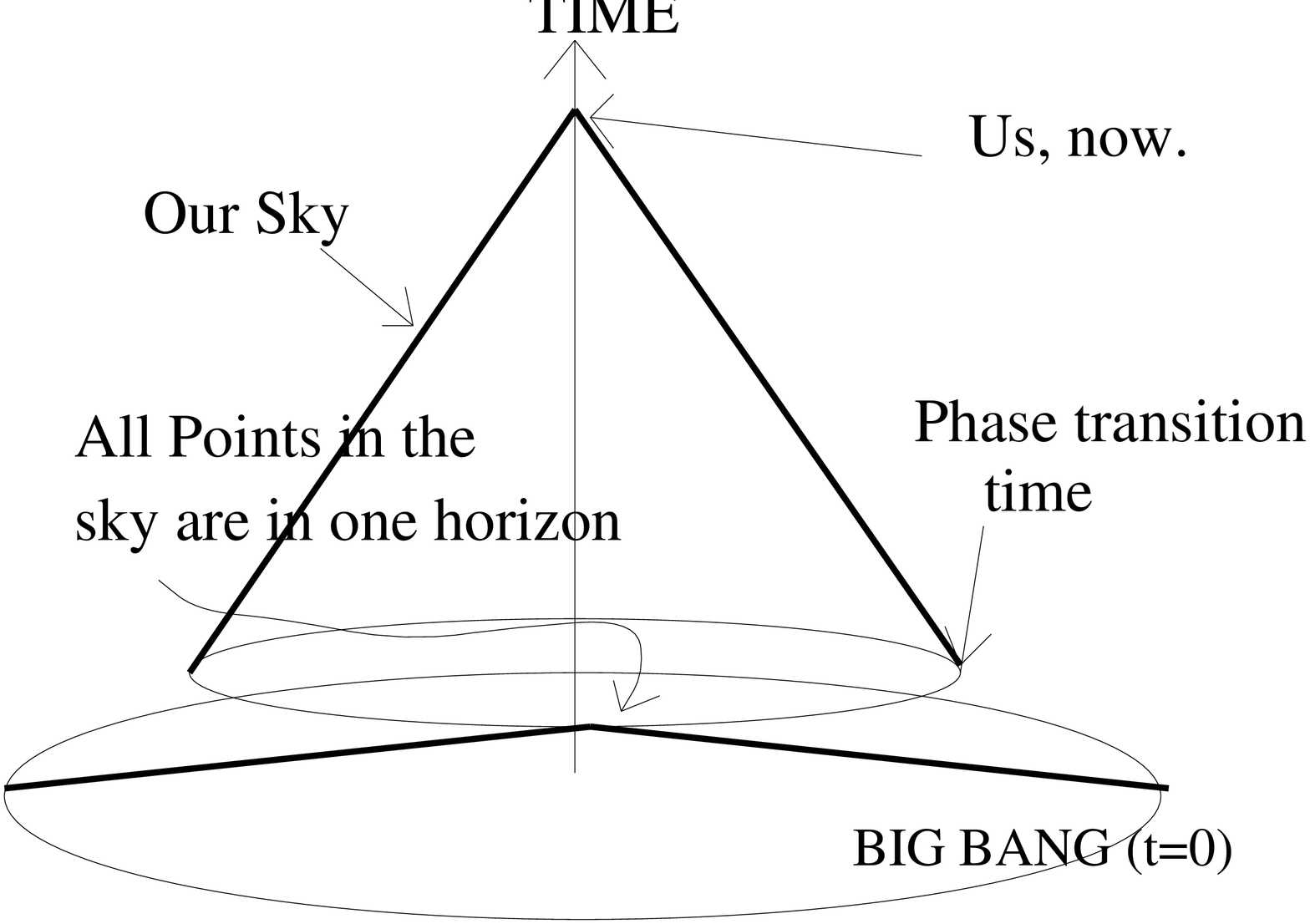,width=6 cm,angle=-90}}
\caption{Diagram showing the horizon structure in a SBB model
in which at time $t_c$ the speed of light changed from $c^-$
to $c^+\ll c^-$. Light travels at $45^\circ$ after $t_c$
but it travels at a much smaller angle with the space axis before
$t_c$. Hence it is possible for the horizon at $t_c$ to be much
larger than the portion of the Universe at $t_c$ intersecting our 
past light cone. All regions in our past have then always been 
in causal contact.}
\label{fig2}
\end{figure}

It is immediately obvious that the horizon problem may be solved
if light
travelled faster in the Early Universe. Suppose there was a ``phase
transition'' at time $t_c$ when the speed of light changed from $c^-$ to
$c^+$. Our past light cone intersects $t=t_c$ at a sphere 
with comoving radius
$r=c^+ (\eta_0-\eta_c)$, where $\eta_0$ and $\eta_c$ are the conformal
times now and at $t_c$. This is as much of the Universe after the
phase transition 
as we can see today. On the other hand the horizon size at $t_c$
has comoving radius $r_h =c^-\eta_c$. If $c^-/c^+\gg\eta_0/\eta_c$,
then $r\ll r_h$, meaning that the whole observable Universe today has
in fact always been in causal contact (see Fig.~\ref{fig2}). 
Some simple  manipulations show
that this requires
\begin{equation}\label{cond1}
\log_{10}{c^-\over c^+}\gg 32 -{1\over 2}\log_{10}z_{eq}+{1\over 2}
\log_{10}{T^+_c\over T^+_P}
\end{equation}
where $z_{eq}$ is the redshift at matter radiation equality, and $T^+_c$
and $T^+_P$ are the Universe and the Planck temperatures after the phase
transition. If $T^+_c\approx  T^+_P$ this implies light travelling more
than 30 orders of magnitude faster before the phase transition. 
It is tempting, for symmetry reasons, simply to postulate that 
$c^-=\infty$.

This is the initial insight: to replace inflationary superluminal
expansion by a varying speed of light. The challenge is now to
set up a consistent VSL theory incorporating this feature, and
possibly solving other cosmological puzzles, such as the
flatness and cosmological constant problems. 

\subsection{The physical meaning of a variable speed of light}\label{mean}
Let us first 
address the question of the meaning of a varying
speed of light. Could such a phenomenon be proved or disproved
by experiment? {\it Physically}
it does not make sense to talk about constancy 
{\it or} variability of any dimensional ``constant''. A measurement of
a dimensional quantity must always represent its ratio to some
standard unit. Hence any measurement is necessarily dimensionless.
It follows that one can only test experimentally the constancy
of dimensionless combinations of constants. 

What would we do therefore if we were to observe changing dimensionless
quantities? Any theory explaining the phenomenon would necessarily have 
to make use of dimensional quantities.  It would a priori be a matter of 
choice, prejudice, or convenience to decide which dimensional quantities 
are variable and which are constant (as we mentioned in the
illustration above).  There would be a kind of equivalence, or duality between
theories based on any two choices as far as dimensionless observations
are concerned. However, the equations for two theories which are 
observationally equivalent, but which have different dimensional
parameters varying, will in general not look the same, and again
simplicity will end up being an important factor in making a choice
between theories.

Consider for instance the recent claim \cite{webb} of experimental evidence 
for a time changing fine structure constant $\alpha=e^2/(4\pi \hbar c)$.
In building a theory which explains a variable $\alpha$
we must make a decision. We could {\it postulate} that electric charge
changes in time, or, say, that $\hbar c$ must change in time. 
Bekenstein \cite{bek2} constructs a theory based on the first alternative. 
He postulates a Lorentz invariant action, which does not conserve 
electric charge. 
Our theory is based on the second choice. We postulate breaking
Lorentz invariance, a changing $\hbar c$,  and consequently 
non-conservation of energy. Any arguments against
the experimental meaning of a changing $c$ can also be directed 
at Bekensteins' changing $e$ 
theory, and such arguments are in both cases meaningless. In both cases 
the choice of a changing dimensional ``constant'' reverts to the postulates
of the theory and is not, a priori, an experimental issue.  

Given a variable $\alpha $, and a VSL theory, it must always be
possible to redefine units so that $c$ and $\hbar $ are constant, and $e$
varies. The two descriptions should be equivalent with respect of
dimensionless quantities. Let us assume that
measurements of intervals of length $dx$, time $dt$, and energy $dE$, are
made in the VSL system of units. In this system $c\propto \hbar \propto 1/{%
\sqrt{\alpha }}$ and $e=e_0$. Now define a new system of units such that the
same measurements in the new system of units lead to results $d{\hat x}$, $d{%
\hat t}$, and $d{\hat E}$ such that 
\begin{eqnarray}
c_0d{\hat x} &=&c\,dx \\
c_0^2d{\hat t} &=&c^2\,dt \\
{\frac{d{\hat E}}{c_0^3}} &=&{\frac{dE}{c^3}}
\end{eqnarray}
where $c_0$ is a constant, to be identified with the fixed speed of light.
These relations fully specify the new system of units. One may then
construct dimensionless ratios in order to identify the constants in the new
system: 
\begin{eqnarray}
{\frac{{\hat c}d{\hat t}}{d{\hat x}}} &=&{\frac{c\,dt}{dx}}  \label{trans0}
\\
{\frac{{\hat \hbar }}{d{\hat E}d{\hat t}}} &=&{\frac \hbar {dE\,dt}} \\
{\frac{{\hat G}d{\hat E}}{d{\hat x}{\hat c}^4}} &=&{\frac{G\,dE}{dx\,c^4}} \\
{\frac{{\hat e}^2}{d{\hat E}d{\hat x}}} &=&{\frac{e_0}{dE\,dx}.}
\end{eqnarray}
From these we find that in the new system of units 
\begin{eqnarray}
{\hat c} &=&c_0 \\
{\hat \hbar } &=&\hbar {\frac{c_0}c}=\hbar _0 \\
{\hat G} &=&G \\
{\hat e} &=&e_0{\frac{c_0}c}\propto {\sqrt{\alpha }.}
\end{eqnarray}
Hence, in the new system, $c$ and $\hbar $ are constants and $e$ varies as $%
\sqrt{\alpha }$. The transformation (\ref{trans0}) can also be defined by 
\begin{eqnarray}
d{\hat x} &=&dx/\epsilon  \label{trans} \\
d{\hat t} &=&dt/\epsilon ^2 \\
d{\hat E} &=&dE\epsilon ^3
\end{eqnarray}
where $\epsilon ={\hat e}/e_0$ is the dielectric constant of the vacuum in
the varying-$e$ system.

\subsection{The foundations of VSL}
In the formulations given to VSL in \cite{AM98}, 
Lorentz invariance is explictly 
broken. In \cite{avel} it has been conjectured that this is a requirement
for the cosmological problems to be solved in these theories. In its current
formulation the foundations of the theory may be summarized in the following
postulates.

{\bf Postulate 1.} {\it A changing $\alpha$ is to be interpreted as
a changing $c$ and $\hbar$ in the ratios $c\propto\hbar\propto
\alpha^{-1/2}$. The coupling $e$ is constant.}

This postulate merely sets up the theoretical interpretation of 
the possible experimental fact that $\alpha$ changes, in terms of variable
dimensional quantities. This is a matter of convention and not
experiment, as much as
a constant $\hbar c$ is a matter of convention. With the above choice
a system of units for mass, length, time, and temperature is unambiguously
defined. 

{\bf Postulate 2.} {\it There is a preferred frame for the laws of physics.
This preferred frame is normally suggested by the symmetries of the 
problem, or by a criterium such as $c=c(t)$.}

If $c$ is variable, Lorentz invariance must be broken. Even if one
writes Lorentz invariant looking expressions these do not transform
covariantly. In general this boils down to the explicit presence of $c$ 
in the operator $\partial_\mu$. Once one admits that Lorentz invariance 
must be explicitly broken then a preferred frame must exist to formulate
the laws of physics. These laws are not invariant under frame 
transformation, and one may expect that a preferred frame exists
where these laws simplify.

{\bf Postulate 3.} {\it In the preferred frame one may obtain the laws of
physics simply by replacing $c$ in the standard (Lorentz invariant) 
action, wherever it occurs,
by a field $c=c(x^\mu)$.}  

This is the principle of minimal coupling. Because the laws of physics
cannot be Lorentz invariant it 
will not hold in every frame. 
Hence the
application of this postulate depends crucially on the previous postulate
supplying us with a favoured frame. This principle may apply in Minkowski
space time electrodynamics, scalar field theory, etc, in which case 
the frame in which $c=c(t)$ is probably the best choice. The cosmological
frame, endowed with the cosmic proper time is probably the best choice
in a cosmological setting.

{\bf Postulate 4} {\it The dynamics of $c$ must be determined by 
an action principle deriving from adding an extra term to the Lagrangian
which is a function of $c$ only.}

 We merely
specify that no fields (including the metric) must be present in this
extra term because we wish minimal coupling to propagate into
the Einstein's equations. On the other hand it is easy to add to 
the $c$ Lagrangian a potential forcing $c$ to do anything we want.
Later we shall explore some possibilities. Here we shall simply consider
the situation is which $c$ is forced to undergo a phase transition
in the early Universe. If the $c$ Lagrangian is Brans Dicke type,
then we have $c\propto a^n$ as in \cite{vsl3}. We shall explore
these scenarios further in Section 6. 

\subsection{Cosmological equations}

In a cosmological setting the postulates proposed imply
that Friedman equations remain valid even when $\dot c\neq 0$:
\begin{eqnarray}
{\left({\dot a\over a}\right)}^2&=&{8\pi G\over 3}\rho -{Kc^2\over a^2}
\label{fried1}\\
{\ddot a\over a}&=&-{4\pi G\over 3}{\left(\rho+3{p\over c^2}\right)}
\label{fried2}
\end{eqnarray}
where, we recall, $\rho c^2$ and $p$ are the energy and 
pressure densities, 
$K=0,\pm 1$ and $G$ the curvature and the gravitational
constants, and the dot denotes a derivative with respect to proper time.
If the Universe is radiation dominated, $p=\rho c^2/3$, and we
have as usual $a\propto t^{1/2}$. 
We have assumed that a frame exists where $c=c(t)$, and identified
this frame with the cosmological frame. 

The assumption that Einstein's equations remain unaffected by 
decelerating light carries with it an important consequence.
Bianchi identities apply to curvature, as a geometrical identity.
These then imply stress energy conservation as an integrability
condition for Einstein's equations. 
If $\dot c\neq 0$,  however,
this integrability condition is not stress energy
conservation. Source terms, proportional to $\dot c/c$,
come about in the conservation equations.

Although this is a general remark we shall be concerned mostly
with violations of energy conservation in a cosmological
setting. Friedman equations can be combined into a 
``conservation equation'' with a source term in 
$\dot c/c$:
\begin{equation}\label{cons1}
\dot\rho+3{\dot a\over a}{\left(\rho+{p\over c^2}\right)}=
{3Kc^2\over 4\pi G a^2}{\dot c\over c}
\end{equation}
In a flat Universe ($K=0$) a changing $c$ does not violate
mass conservation. Energy, on the other hand, 
is proportional to $c^2$. If,
however, $K\neq 0$ not even mass is conserved.

\section{The cosmological problems and VSL}
Even though it is obvious that a varying speed of light theory
solves the horizon problem, it is far from obvious that other
cosmological porblems may be solved as well. Here we concentrate
on the flatness, cosmological constant, and homogeneity problems.

\subsection{The flatness puzzle}
The flatness puzzle can be illustrated as follows.
Let $\rho_c$ be the critical density of the Universe:
\begin{equation}
\rho_c={3\over8\pi G}{\left(\dot a\over a\right)}^2
\end{equation}
that is, the mass density corresponding to $K=0$
for a given value of $\dot a/a$. Let us define
$\epsilon=\Omega-1$ with $\Omega=\rho/\rho_c$. Then  
\begin{equation}
\dot\epsilon=(1+\epsilon){\left({\dot\rho\over\rho}-
{\dot\rho_c\over\rho_c}
\right)}
\end{equation}
If $p=w\rho c^2$ (with $\dot w=0$), using 
Eqns.(\ref{fried1}), (\ref{fried2}), and 
(\ref{cons1}) we have:
\begin{eqnarray}
{\dot\rho\over \rho}&=&-3{\dot a\over a}(1+w)+
2{\dot c\over c}{\epsilon\over 1+\epsilon}\\
{\dot\rho_c\over \rho_c}&=&-{\dot a\over a}(2+(1+\epsilon)(1+3w))
\end{eqnarray}
and so
\begin{equation}\label{epsiloneq}
\dot\epsilon=(1+\epsilon)\epsilon {\dot a\over a} 
{\left(1+3w\right)}+2{\dot c\over c}\epsilon
\end{equation}
In the SBB $\epsilon$ grows like $a^2$ in the radiation era, 
and like $a$ in the matter era, leading to a total growth by 
32 orders of magnitude since the Planck epoch. The observational 
fact that $\epsilon$ can at most be of order 1
nowadays requires that either $\epsilon=0$
strictly, or an amazing fine tuning must have existed in the initial
conditions ($\epsilon<10^{-32}$ at $t=t_P$). This is the flatness puzzle.

The $\epsilon=0$ solution is in fact unstable for any matter 
field satisfying the strong energy condition $1+3w>0$. Inflation
solves the flatness problem with an inflaton field which satisfies
$1+3w<0$. For such a field $\epsilon$ is driven towards
zero instead of away from it. Thus inflation can solve the
flatness puzzle.

As Eqn.~\ref{epsiloneq} shows a decreasing speed of light 
($\dot c/c<0$) would also drive $\epsilon$ to 0. If the speed 
of light changes in a sharp phase transition, with $|\dot c/c|\gg
\dot a/a$, we can neglect the expansion terms in 
Eqn.~\ref{epsiloneq}. Then $\dot\epsilon/\epsilon=2\dot c/c$ so
that $\epsilon\propto c^2$. A short calculation shows that the 
condition (\ref{cond1}) also ensures 
that $\epsilon\ll 1$ nowadays, if $\epsilon\approx 1$ before the
transition. 

The instability of the $K\neq 0$ Universes while $\dot c/c<0$ can be
expected simply from inspection of the non conservation equation
Eq.~(\ref{cons1}). Indeed if $\rho$ is above its critical value,
then $K=1$, and Eq.~(\ref{cons1}) tells us that mass is taken out 
of the Universe. If $\rho<\rho_c$, then $K=-1$, and then mass is produced.
Either way the mass density is pushed towards its critical value
$\rho_c$. In contrast with the Big Bang model, during a period 
with $\dot c/c<0$ only the $K=0$ Universe is stable.

We have assumed in the previous discussion that we are close,
but not fine-tuned, to flatness before the transition. 
It is curious to note that this need not be the case. 
Suppose instead that the Universe acquires ``natural initial
conditions'' (eg. $\epsilon\approx 1$) well
before the phase transition occurs. If such Universes 
are closed they recollapse before the transition. If they are
open, then they approach $\epsilon=-1$.  This is the Milne Universe,
which is our case (constant $G$) may be seen as Minkowski space-time. 
Such a curvature dominated Universe is essentially empty, and a coordinate 
transformation can transform it into Minkowski space-time. Inflation
cannot save these empty Universes, as can be seen from Eqn.~\ref{epsiloneq}.
Indeed even if $1+3w<0$ the first term will be
negligible if $\epsilon\approx-1$. This is not true for VSL: the
second term will still push an $\epsilon=-1$ Universe towards
$\epsilon=0$.

Heuristically this results from the fact that the violations of 
energy conservation responsible for pushing the Universe towards 
flatness do not depend on there being any matter in the Universe.
This can be seen from inspection of Eqn.~(\ref{cons1}).

In this type of scenario it does not matter how far before
the transition the ``initial conditions'' are imposed. We 
end up with a chaotic scenario in which Darwinian selection gets rid
of all the closed Universes. The open Universes become empty and cold.
In the winter of these Universes a phase transition
in $c$ occurs, producing matter, and leaving the Universe
very fine tuned, indeed as an Einstein deSitter Universe (EDSU). 
This point seems to have been missed in \cite{cole}.

\subsection{The cosmological constant problem}
There are two types of cosmological constant problems, and
we wish to start our discussion by differentiating them.
Let us write the action as:
\begin{equation}
S=\int dx^4 \sqrt{-g}{\left( {c^4 (R+2\Lambda_1)\over 16\pi G}
 +{\cal L}_M + {\cal L}_{\Lambda_2}\right)}
\end{equation}
where ${\cal L}_M$ is the matter fields Lagragian. 
The term in $\Lambda_1$ is a geometrical cosmological constant,
as first introduced by Einstein. The term in $\Lambda_2$ represents
the vacuum energy density of the quantum fields \cite{steve}.
Both tend to dominate the energy density of the Universe,
leading to the so-called cosmological constant problem.
However they represent two rather different problems.
We shall attempt to solve the problem associated with 
the first, not the second, term.
Ususally one hopes that the second term will be cancelled by an
additional couter-term in the Lagrangian. In the rest of
this paper it is the geometrical cosmological constant
that is under scrutiny.

If the cosmological constant $\Lambda\neq 0$ then the
argument in the previous section
still applies, with $\rho=\rho_m+\rho_\Lambda$,
where $\rho_m$ is the mass density in normal matter, and
\begin{equation}\label{enerlamb}
\rho_\Lambda={\Lambda c^2\over 8\pi G}
\end{equation} 
is the mass density in the cosmological constant.
One still predicts $\Omega_m+\Omega_\Lambda=1$, with
$\Omega_m=\rho_m/\rho_c$ and $\Omega_\Lambda=\rho_\Lambda/\rho_c$.
However now we also have
\begin{equation}\label{dotLm}
\dot\rho_m+3{\dot a\over a}{\left(\rho_m+{p_m\over c^2}
\right)}=-\dot\rho_\Lambda+
{3K c^2\over 4\pi G a^2}{\dot c\over c}
\end{equation} 
If $\Lambda$ is indeed a constant then from Eq.~(\ref{enerlamb})
\begin{equation}\label{dotL}
{\dot \rho_\Lambda\over \rho_\Lambda}=2{\dot c\over c} 
\end{equation} 
If we define $\epsilon_\Lambda=\rho_\Lambda/\rho_m$
we then find, after some straightforward algebra, that
\begin{equation}\label{epslab}
\dot \epsilon_\Lambda =\epsilon_\Lambda{\left(
3{\dot a\over a}(1+w)+2{\dot c\over c}{1+\epsilon_\Lambda
\over 1+\epsilon}\right)}
\end{equation} 
Thus, in the SBB model, 
$\epsilon_\Lambda$  increases like $a^4$ in the radiation era, 
like $a^3$ in the matter era,
leading to a total growth by 64 orders of magnitude since the Planck
epoch. 
Again it is puzzling that $\epsilon_\Lambda$ is observationally
known to be at most of order 1
nowadays. We have to face another fine tuning problem in the SBB
model: the cosmological constant problem.

If $\dot c=0$ the solution $\epsilon_\Lambda=0$
is in fact unstable for any $w>-1$. Hence violating the  strong
energy condition $1+3w>0$ would not solve this problem.
Even in the limiting case $w=-1$ the solution 
$\epsilon_\Lambda=0$ is not an attractor: $\epsilon_\Lambda$
would merely remain constant during inflation, then starting to
grow like $a^4$ after inflation.
Therefore inflation cannot ``explain'' the small value
of $\epsilon_{\Lambda}$, as it can with $\epsilon$,
unless one violates the dominant energy condition
$w\ge -1$. 

However, as Eqn.~(\ref{epslab}) shows, 
a period with $\dot c/ c \ll 0$ would drive $ \epsilon_\Lambda$
to zero.  If the speed of light changes suddenly ($|\dot c/c|
\gg \dot a/a$) then we can neglect terms in $\dot a/a$, and so
\begin{equation}
{\dot \epsilon_\Lambda\over \epsilon_\Lambda(1+\epsilon_\Lambda)}
=2{\dot c\over c}{1\over 1+\epsilon}
\end{equation} 
which when combined with $\dot\epsilon/\epsilon=2\dot c/c$
leads to
\begin{equation}
{\epsilon_\Lambda\over 1+\epsilon_\Lambda}
\propto {\epsilon \over 1+\epsilon}
\end{equation} 
The exact constraint on the required 
change in $c$ depends on the initial conditions
in $\epsilon$ and $\epsilon_\Lambda$. In any case once both
$\epsilon\approx 1$ and $\epsilon_\Lambda\approx 1$ we have 
$\epsilon_\Lambda\propto c^2$. Then we can solve the 
cosmological constant problem in a sudden phase transition 
in which
\begin{equation}\label{cond2}
\log_{10}{c^-\over c^+}\gg 64 -{1\over 2}\log_{10}z_{eq}+2\log_{10}
{T^+_c\over T^+_P}
\end{equation}
This condition is considerably more restrictive than (\ref{cond1}),
and means a change in $c$ by more than 60 orders of magnitude,
if $T^+_c\approx  T^+_P$.
Note that once again a period with $\dot G/G$ would not solve
the cosmological constant problem.

As in the case of the flatness problem we do not need to impose
``natural initial conditions'' ($\epsilon_\Lambda\approx 1$)
just before the transition. These could have existed any time
before the transition, and the argument would still go through,
albeit with a rather different overall picture for the history of the
Universe.

If $\epsilon_\Lambda\approx 1$ well before the transition, then
the Universe soon becomes dominated by the cosmological constant.
We have inflation! The curvature and matter will be inflated away.
We end up in a de-Sitter Universe. When the transition is about to occur
it finds a flat Universe ($\epsilon=0$), with no matter ($\rho_m=0$),
and with a cosmological constant. If we rewrite Eqn.(\ref{epslab})
in terms of $\epsilon_m=\rho_m/\rho_\Lambda$, for $\epsilon=0$
and $|\dot c/c|\gg \dot a /a$, we have 
$\dot \epsilon_m=-2(\dot c/ c)(1+\epsilon_m)$. Integrating
leads to $1+\epsilon_m\propto c^{-2}$.
We conclude that we do not need the presence of any matter in the Universe
for a VSL transition to convert a cosmological
constant dominated Universe into a EDSU Universe full of ordinary
matter. This can be seen
from Eqns.~(\ref{dotLm})-(\ref{dotL}). A sharp decline in $c$ will
always discharge any vacuum energy density into ordinary matter.

We stress the curious point that 
in this type of scenario the flatness problem is not
solved by VSL, but rather by the period of inflation
preceding VSL.

\subsection{The homogeneity of the Universe}\label{homo}
Solving the horizon problem by no means guarantees solving 
the homogeneity problem, that is, the uncanny homogeneity of 
the currently observed Universe
across many regions which have apparently been causally disconnected.
Although solving the horizon problem is a necessary condition for solving
the homogeneity problem, in a generic inflationary model solving the 
first causes serious 
problems in solving the latter.  Early causal contact between
the entire observed Universe allows
equilibration processes to homogenize the whole observed
Universe.  It is crucial to the inflation picture that before
inflation the observable universe in well inside the Jeans length,
and thus equilibrates toward a homogeneous state.
However no such process is perfect, and small density
fluctuations tend to be left outside the Hubble radius, 
once the Universe resumes its
standard Big Bang course. These fluctuations then grow like $a^2$
during the radiation era, like $a$ during the matter era, usually entailing
a very inhomogeneous Universe nowadays. This is a common flaw in
early inflationary models \cite{gupi} which requires additional
fine-tuning to resolve.

In order to address this problem in VSL a rather complex calculation
needs to be carried out. It consists in taking the perturbed
Einstein's equations assuming a changing $c$, and from them
deriving the equations of motions for the matter perturbation.
The basic result is that the comoving
density contrast $\Delta$ and gauge-invariant velocity $v$
are subject to the equations:
\begin{eqnarray}
\Delta'-{\left(3w{a'\over a}+{c'\over c}\right)}\Delta&=&
-(1+w)kv-2{a'\over a}w\Pi_T\label{delcdotm}\\
v'+{\left({a'\over a}-2{c'\over c}\right)}v&=&{\left(
{c_s^2 k\over 1+w} -{3\over 2k}
{a'\over a}{\left({a'\over a}+{c'\over c}\right)}
\right)}\Delta \nonumber \\
+{kc^2w\over  1+w}\Gamma-
&kc&{\left({2/3\over 1+w}+{3\over k^2c^2}
{\left(a'\over a\right)}^2\right)}w\Pi_T\label{vcdotm}
\end{eqnarray}
where $k$ is the wave vector of the fluctuations,
and $\Gamma$ is the entropy production rate, $\Pi_T$
the anisotropic stress, and $c_s$ the speed of sound given by:
\begin{equation}\label{cs}
c^2_s={p'\over \rho '}=wc^2{\left(1-{2\over 3}{1\over 1+w}
{c'\over c}{a\over a'}\right)}
\end{equation}

In the case of a sudden phase transition Eqn.~({\ref{delcdotm})
shows us that $\Delta\propto c$, regardless of the chosen 
equations of state for $\Gamma$ and $\Pi_T$. Hence 
\begin{equation}
{\Delta^+\over\Delta^-}={c^+\over c^-}
\end{equation}
meaning a suppression of any fluctuations before the phase
transition by more than a factor of $10^{-60}$ if condition
(\ref{cond2}) is satisfied.
The suppression of fluctuations induced by a sudden phase transition
in $c$ can be intuitively understood in the same fashion as the solution to
the flatness problem. Mass conservation violation
ensures that only a Universe at critical mass density is stable,
if $\dot c/c\ll 0$. But this process occurs locally, so
after the phase transition the Universe should be left
at critical density {\it locally}. Hence the suppression of 
density fluctuations.

We next need to know what are the initial conditions for $\Delta$ and
$v$. Suppose that at some very early time $t_i$
one has $\dot c/c= 0$ and the whole observable Universe
nowadays is inside the Jeans length: $\eta_0\ll c_i\eta_i/{\sqrt 3}$.
The latter condition is enforced as a byproduct of solving the horizon
problem. The whole observable Universe nowadays is then
initially in a thermal state. What is more each portion of the Universe
can be described by the canonical ensemble and so the Universe
is homogeneous apart from thermal fluctuations \cite{Peebles}.
These are characterized by the mass fluctuation 
\begin{equation}
\sigma^{2}_M={{\langle\delta M ^2
\rangle}\over{\langle M\rangle}^2}={4k_b T_i\over M c_i^2}
\end{equation}
Converted into a power spectrum for $\Delta$ this is a 
white noise spectrum with amplitude
\begin{equation}\label{pdelta}
P_\Delta(k)={\langle |\Delta(k)^2|\rangle}\propto 
{4k_bT_i\over \rho_ic_i^2}
\end{equation}

What happens to a thermal distribution, its temperature, and its
fluctuations, while $c$ is changing?
In thermal equilibrium the distribution function of particle energies
is the Planck distribution $P(E)=1/(e^{E/k_bT}-1)$, where $T$
is the temperature. 
When one integrates over the whole phase space, one obtains
the bulk energy density $\rho c^2\propto (k_b T)^4/(\hbar c)^3$.
Let us now consider the time when the Universe has already 
flattened out sufficiently for mass to be approximately 
conserved. To define the situation more completely, we 
make two additional microphysical assumptions.
Firstly, let mass be conserved also for individual quantum particles,
so that their energies scale like $E\propto c^2$. 
Secondly,  we assume particles' wavelengths do not change with $c$. 
If homogeneity is preserved, indeed the wavelength is an 
adiabatic invariant, fixed by a set of quantum numbers, 
eg: $\lambda =L/n$ for a particle in a box of size $L$.
 
Under the first of these assumptions a Planckian distribution with
temperature $T$ remains Planckian, but $T\propto c^2$.
Under the second assumption, we have $\lambda=2\pi \hbar c/E$, 
and so $\hbar/ c$ should remain constant. Therefore the phase space 
structure is changed so that, without particle production, one still
has $\rho c^2\propto (k_b T)^4/(\hbar c)^3$, with $T\propto c^2$.
A black body therefore remains a black body,
with a temperature $T\propto c^2$. If we combine this effect 
with expansion, with the aid of Eqn.~(\ref{cons1}) we have
\begin{equation}\label{temp}
\dot T + T{\left({\dot a\over a}-2{\dot c\over c}\right)}=0
\end{equation}
We can then integrate this equation through the epoch when
$c$ is changing to find the temperature $T_i$ of the initial
state. This fully fixes the initial conditions for scalar
fluctuations, by means of (\ref{pdelta}). 

In the case of a sudden phase transition we have $T^+=
T^- c^{2+}/c^{2-}$, and so 
\begin{equation}
\sigma_M^{2-}={4k_b T^-\over M c^{2-}}={4k_b T^+\over M c^{2+}}
\end{equation}
or 
\begin{equation}
\Delta^-(k)^{2}\approx {4k_bT^+\over \rho^+ c^{2+}}
\end{equation}
but since $\Delta\propto c$ we have 
\begin{equation}
\Delta^+(k)\approx {\sqrt {4k_bT^+\over \rho^+ c^{2+}}}{c^+\over c^-}
\end{equation}
Even if $T^+=T_P^+=10^{19}Gev$ these fluctuations would still be 
negligible nowadays. Therefore although the Universe ends up in a 
thermal state after the phase transition, its thermal fluctuations,
associated with the canonical ensemble, are strongly suppressed.

\section{A possible experimental test of VSL}
VSL, like inflation, may always evade conflict with experiment. In the
phase transition scenario of Albrecht and Magueijo, changes in $c$
occur only in the very Early Universe. They are beyond
a chance of contradiction with experiment.

This situation is not entirely satisfactory. A theory which cannot
be falsified, cannot also be true! In this section we show how it
is possible to put VSl to the test. We follow recent work by
Barrow and Magueijo \cite{sn}. In this work, the authors propose a 
dynamical varying $c$, that is a theory in which the changes in
$c$ can be predicted from the matter distribution, rather than 
being just postulated. The model then allows for a joint explanation
of two recent observations. 

One is the Supernovae 
Cosmology Project and the High-z Supernova Search
(\cite{super,super1,super2,super3}). These extended the reach of the 
Hubble diagram to high redshift and provided new 
evidence that the expansion of the universe is accelerating. 
This {\it may} imply 
that there exists a significant positive cosmological constant, $\Lambda $. 
The other observation \cite{webb} was already discussed above:
recent evidence for a redshift dependence of the fine structure
constant. 

In this section we shall first reinterpret the Hubble diagram
in VSL theories, following which we shall show how a dynamical
model for a changing $c$ may explain these two observations.

\subsection{The VSL Hubble diagram} 
 
\label{hubble} The Hubble diagram is a plot of luminosity distance against 
redshift. The purpose is to map the expansion factor $a(t)$, where $t$ is 
the comoving proper time. Redshifts provide a measurement of $a$ at the time 
of emission. If the objects under observation are ``standard candles'' (as 
Type Ia supernovae are assumed to be), their apparent brightness gives their 
(luminosity) distance, which, if we know $c$, tells us their age. By looking 
at progressively more distant objects we can therefore map the curve $a(t)$. 
 
We now examine how this construction is affected by a changing $c$. In  
\cite{AM98} we showed that $E\propto c^2$ for photons in free flight.  
We also showed that quantum mechanics remains unaffected by a changing 
$c$ if $\hbar\propto c$ (in the sense that quantum numbers are adiabatic 
invariants). If 
$\hbar \propto c$, the Rydberg energy scales like $c^2$:  
\begin{equation} 
E_R={\frac{m_ec^4e}{2\hbar ^2}}^4\ \propto c^2 
\end{equation} 
Hence all absorption lines, ignoring the fine structure, scale like $c^2$. 
When we compare lines from near and far systems we should therefore see no 
effects due to a varying $c$; the redshift $z$ is still  
\begin{equation}\label{z}  
1+z_e=a_o/a_e  
\end{equation} 
where $o$ and $e$ label epochs of observation and emission.
 
In order to examine luminosity distances, we need to reassess the concept of 
standard candles. For simplicity let us first treat them as black bodies. 
Then their temperature 
scales as $T\propto c^2$ (\cite{AM98}), their energy density scales as $\rho 
\propto T^4/(\hbar c)^3\propto c^2$, and their emission power as $P=\rho 
/c\propto c$, implying that standard candles are brighter in the early 
universe if $\dot c<0$. However, the power emitted by these candles, in free 
flight, scales like $c$; each photon's energy scales like $c^2$, its speed 
like $c$, and therefore its energy flux like $c$. The received flux, as 
a function of $c$, therefore scales like: 
\begin{equation} 
P_r={\frac{P_ec^2}{4\pi r^2c}}\propto c 
\end{equation} 
where $r$ is the conformal distance to the emitting object, and the  
subscripts $r$ and $e$ label received and emitted.  
In an expanding universe we therefore still have  
\begin{equation} 
P_r={\frac{P_{e0}}{4\pi r^2a_0^2}}{\left( \frac a{a_o}\right) }^2, 
\label{lum} 
\end{equation} 
where $P_{e0}$ is the emitting power of standard candles today. 
Notice that the above argument is still valid if the candles 
are not black bodies; it depends only on the scaling properties 
of emitted and received power. 
 
We can now set up the Hubble diagram. Consider the Taylor expansion  
\begin{equation} 
a(t)=a_0[1+H_0(t-t_0)-{\frac 12}q_0H_0^2(t-t_0)^2+...]  \label{expansion} 
\end{equation} 
where $H_0=\dot a_0/ a_0$ is the Hubble constant, and 
$q_0=-\ddot a_0 a_0/ \dot a_0^2$ is the decceleration parameter. 
Hence 
\begin{equation} 
z=H_0(t_0-t)+(1+q_0/2)H_0^2(t-t_0)^2+...., 
\end{equation} 
or  
\begin{equation} 
t_0-t={\frac 1{H_0}}[z-(1+q_0/2)z^2+...].  \label{t-t0} 
\end{equation} 
From (\ref{lum}) we find that the luminosity distance $d_L$ is  
\begin{equation} 
d_L={\left( \frac{P_{e0}}{4\pi P_0}\right) }^{1/2}=a_0^2{\frac ra}%
=a_0r(1+z_e).
\end{equation} 
The conformal distance to the emitting object is given by  
\begin{equation} 
r=\int_t^{t_0}{\frac{c({t})d{t}}{a({t})}.} 
\end{equation} 
From (\ref{expansion}) we have that  
\begin{equation} 
r=c_0[(t_0-t)+{\frac{1-n}2}H_0(t_0-t)^2+...] 
\end{equation} 
where we have assumed that locally $c=c_0a^n$ (that is  
$c=c_0[1+nH_0(t-t_0)+...]$). Substituting (\ref{t-t0}) we 
finally have :  
\begin{equation} 
d_L={\frac{c_0}{H_0}}[z+{\frac 12}(1-(q_0+n))z^2+...] 
\end{equation} 
We see that besides the direct effects of VSL upon the expansion rate of the 
universe, it also induces an effective{\it \ acceleration} in the Hubble 
diagram as an ``optical illusion'' (we are assuming that $c$ decreases in 
time: $n<0$). This is easy to understand. We have seen that VSL introduces 
no intrinsic effects in the redshifting spectral line 
or in the dimming of standard 
candles with distance and expansion.  
The only effect VSL induces on the 
construction of the Hubble diagram is that for the same redshift (that is, 
the same distance into the past) objects are farther away from us because 
light travelled faster in the past. But an excess luminosity distance, for 
the same redshift, is precisely the hallmark of cosmological acceleration. 
However, we need to consider the other experimental input to our work: the  
\cite{webb} results. By measuring the fine structure in absorption systems 
at redshifts $z\sim O(1)$ we can also map the curve $c(t)$. Since $%
c=c_0[1+nH_0(t-t_0)+...]$ we have $c=c_0[1-nz+...]$, and so to first order $%
\alpha =\alpha _0[1+2nz+...]$. However, the results presented in  \cite 
{webb} show that $n$ is at most of order $10^{-5}.$ This means that the 
direct effects of varying $c$ permitted by the quasar absorption system 
observations are far too small to explain the observed acceleration. We need 
to look at a fully self-consistent generalisation of general relativity 
containing the scope for varying $c.$ 
 
\subsection{The model} 
 
\label{model} We start with some general properties of the dynamics of $c$.
Drawing inspiration from dilaton 
theories (like Brans-Dicke gravity) we take  
\[ 
\psi =\log (c/c_0)  
\] 
as the dynamical field associated with $c$. 
 Indeed, powers of $c$ appear in all coupling constants, which in turn can be 
written as $e^\phi $, where $\phi $ is the dilaton. We then endow $\psi $ 
with a dynamics similar to the dilaton. The left-hand side for the $\psi $ 
equation for a homogeneous field in an expanding universe should be ${\ddot 
\psi }+3{\frac{\dot a}a}{\dot \psi }$. This structure ensures that the 
propagation equation for $\psi $ is second-order and hyperbolic. However, 
since VSL breaks Lorentz invariance one need not choose (as in Brans-Dicke) 
the source term to be $\rho -3p$, where $\rho $ and $p$ are the energy 
density and pressure of matter respectively. A possibility is
\begin{equation}\label{dyn2} 
{\ddot \psi}+3{\dot a\over a}\dot \psi=4\pi G \omega\rho 
\end{equation} 
We could have added pressure to the source  and terms in the curvature. 
An interesting alternative is  
\begin{equation}\label{dyn1} 
{\ddot \psi}+3{\dot a\over a}\dot \psi=4\pi G \omega 
(\rho-p/c^2)-{2Kc^2\omega\over a^2} 
\end{equation} 
These dynamical equations, when combined with Friedmann equations,
lead to a ``Machian'' scenario, in which $c=c_0a^n$. 
If we consider the dynamical equation (\ref{dyn1}), then
 with $n=\omega$, regardless of the equation  
of state, or amount of curvature domination. This is the scenario initially 
envisaged by Barrow \cite{barrow}. We now see that the  
coupling constant $\omega$,  
describing the ability of matter to drive changes in the speed of light, 
is also what determines the exponent $n$ in the Machian solution. 
The flatness problem is solved if: 
\begin{equation} 
\omega<(1-3\gamma/2) 
\end{equation} 
Effectively this means $\omega<-1$, since the flatness problem 
should be solved in the radiation epoch. This is a self regulating 
scenario, in which curvature drives changes in the speed of light, 
which in turn suppress curvature.  
If we take Eqn.(\ref{dyn2}) then if the curvature term is 
subdominant one has  
\begin{equation} 
n=\omega/(2-\gamma) 
\end{equation} 
that is $n=3/2\omega$ in the radiation epoch, and $n=\omega$  
in the matter epoch. The flatness problem is solved if 
\begin{equation} 
\omega<(2-\gamma)(1-3\gamma/2) 
\end{equation} 
Effectively this means $\omega<-2/3$, since the flatness problem 
should be solved in the radiation epoch. Assuming that curvature  
has already been dominated is natural in these scenarios, as there  
is no Planck time. Hence there is no natural 
time when we may give matter and curvature initial values of the same 
order of magnitude. It is sufficient for flatness to be an attractor. 

In order to explain the Supernovae results, however, the best dynamical
equation is 
\begin{equation} 
{\ddot \psi }+3{\frac{\dot a}a}{\dot \psi }=4\pi G\omega {\frac p{c^2},} 
\label{sor} 
\end{equation} 
where $p$ is the total pressure of the matter fields and $\omega $ is a 
coupling constant (distinct from the Brans Dicke coupling constant).  
The full self-consistent system of equations in a 
matter-plus-radiation universe containing a cosmological constant stress 
 
\[ 
\rho _{\Lambda  
}=\frac{\Lambda c^2}{8\pi G},  
\] 
is therefore  
\begin{eqnarray} 
\ddot \psi +3{\frac{\dot a}a}\dot \psi &=&4\pi G\omega {\frac{\rho _\gamma }{%
3\label{totaleqns1}},} \\ 
\dot \rho _\gamma +4{\frac{\dot a}a}\rho _\gamma &=&-2\rho _\Lambda \dot 
\psi ,  \label{totaleqns2} \\ 
\dot \rho _\Lambda &=&2\rho _\Lambda \dot \psi ,  \label{totaleqns3} \\ 
\dot \rho _m+3{\frac{\dot a}a}\rho _m &=&0,  \label{totaleqns4} \\ 
{\left( \frac{\dot a}a\right) }^2 &=&{\frac{8\pi G}3}(\rho _m+\rho _\gamma 
+\rho _\Lambda ),  \label{totaleqns5} 
\end{eqnarray} 
where subscripts $\gamma $ and $m$ denote radiation and matter respectively. 
We have assumed that the sink term (\ref{totaleqns3}) is reflected in 
a source term in (\ref{totaleqns2}) (and not in (\ref{totaleqns4})). 
This is due to the fact that this term is only significant very early on, 
when even massive particles behave like radiation.  
We have ignored curvature terms because in the quasi-lambda dominated 
solutions we are about to explore we know that these are smaller than $%
\rho _{\Lambda  
}$, (\cite{vsl3}).   
Here, in complete contrast to 
Brans-Dicke theory, the field $\psi $ is only driven by radiation pressure 
in the dust-dominated era. In other words, only conformally invariant 
forms of matter couple to the field $\psi$.  
 
\subsection{Solutions to the model}

In a radiation-dominated universe the behaviour of this system changes at 
the critical value $\omega =-4$. For $\omega <-4$ we reach a flat $\Lambda 
=0 $ attractor as $t\rightarrow \infty $. For $-4<\omega <0$ we have 
attractors for which $\rho _\Lambda $ and $\rho _\gamma $ maintain a 
constant ratio (see \cite{vsl3}). In Fig.~\ref{fig1b} we plot a numerical 
solution to 
this system, with $\omega =-4.4$ (a $10\%$ tuning below the critical value $%
\omega =-4$) and $n=-2.2$ during the radiation epoch. As expected from \cite 
{vsl3}, this forces $\Omega _\Lambda $ to drop to zero, while the expansion 
factor acquires a radiation-dominated form, with $a\propto t^{1/2}$. By the 
time the matter-dominated epoch is reached, $\Omega _\Lambda $ is of order $%
10^{-12}$. During the matter epoch, the source term for $\psi $ disappears 
in eq. (\ref{sor}), $n$ starts to approach zero, $\Omega _\Lambda $ starts 
to increase, and the expansion factor takes on the $a\propto t^{2/3}$ 
dependence of a matter-dominated universe. A few expansion times into the 
matter epoch, $\Omega _\Lambda $ becomes of order 1 and the universe begins 
accelerating. By the time this happens $n$ is of order $10^{-5}$, in 
agreement with the expectations of \cite{webb}. This type of behaviour 
can be achieved generically, for different initial conditions, with a tuning 
of $\omega $ that never needs to be finer than a few percent. 
 
We can provide an approximate argument explaining why this theory should 
display this type of behaviour and why we need so little fine tuning of $%
\omega $ to explain the supernovae experiments. If we neglect changes in $c$ 
after matter-radiation equality, $t_{eq}$, we are going to require  
\begin{equation} 
{\frac{\rho _\Lambda (t_{eq})}{\rho (t_{eq})}}\approx z_{eq}^{-3}\sim 
10^{-12}. 
\end{equation} 
Let $c=c_0a^{n(t)}$, with $n=-2-\delta $, and $n=\omega /2$ during the 
radiation epoch. We can integrate the conservation equations to give  
\begin{equation} 
{\frac \rho {\rho _\Lambda }}={\frac A{a^4\rho _\Lambda }}-{\frac n{n+2},} 
\end{equation} 
with $A$ constant, from which it follows that  
\begin{equation} 
{\frac \rho {\rho _\Lambda }}={\frac 2\delta }{\left[ {\left( 1+{\frac 
\delta 2}{\frac{\rho _i}{\rho _{\Lambda i}}}\right) }{\left( \frac 
a{a_i}\right) }^{2\delta }-1\right] .} 
\end{equation} 
We see that assymptotically $\rho/\rho_\Lambda$ grows to infinity, 
if $\delta >0$ (the flat $\Lambda=0$ attractor of \cite{vsl3}). 
However the growth is very slow even if $\delta$ is not very small. 
Our theory displays very long transients, and a very slow convergence 
to its attractor, a property similar to quintessence models 
(\cite{quint}).  
It is therefore possible to achieve $\rho _\Lambda /\rho \sim 10^{-12}$ at 
the end of the radiation epoch, with $\delta $ chosen to be of order $0.1$. 
 \begin{figure}
\centerline{{\psfig{file=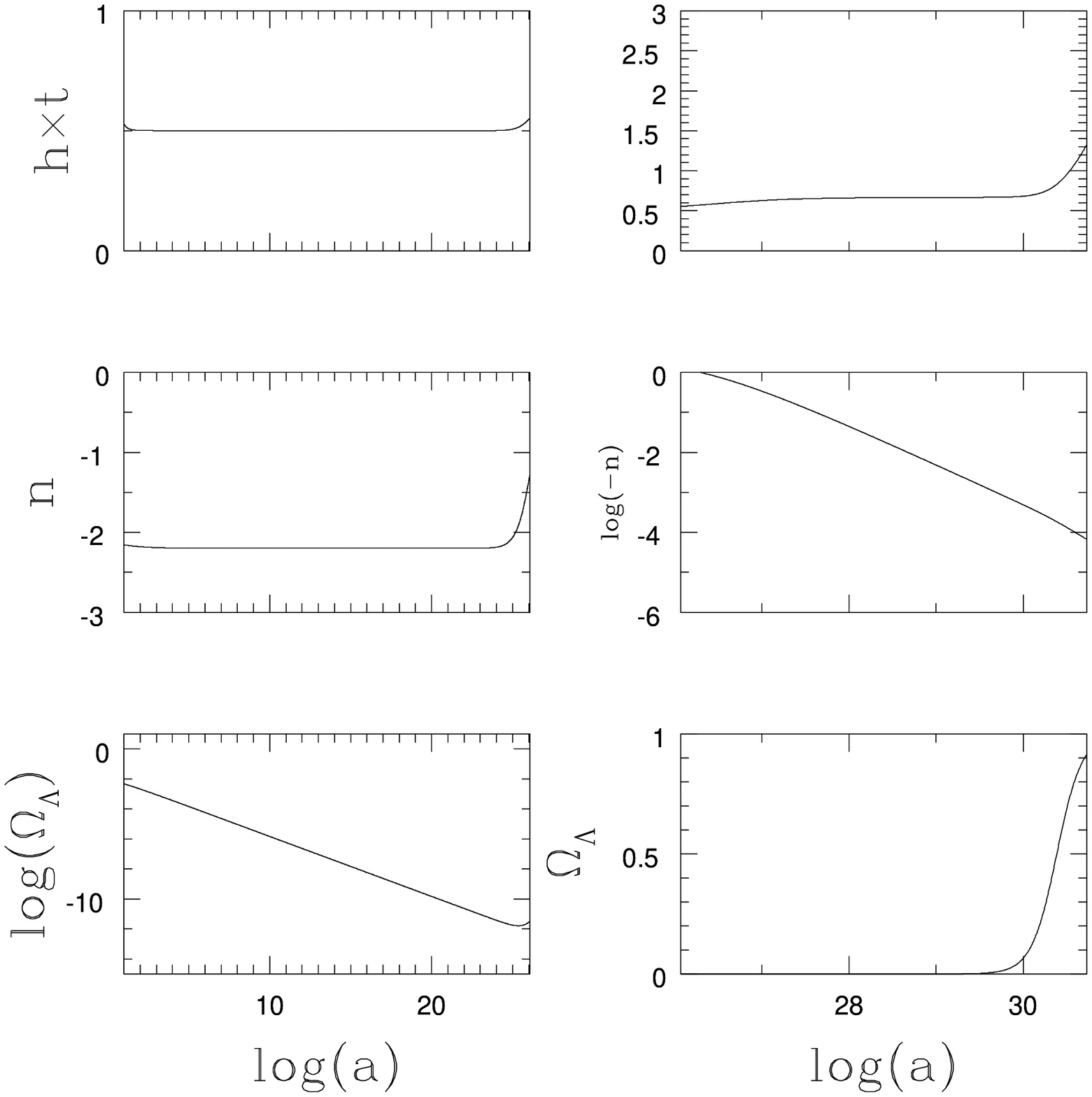,width=12 cm,angle=0}}} 
\caption{Evolution of $t\times h$ (where $h= {\frac{\dot a}{a}}$), 
$n={\frac{\dot c/c}{\dot 
a /a}}$, and $\Omega_\Lambda$ in log or linear plots as appropriate. The 
panels on the left (right) describe the radiation (matter) dominated epoch. 
We have taken $\omega=-4.4$ (a modest $10\%$ tuning over the critical value
$\omega=-4$). In the radiation epoch $n=-2.2$, 
$\Omega_\Lambda$ slowly drops to zero, and 
the expansion factor has the usual dependence $a\propto t^{1/2}$.  
As the Universe enters the matter epoch $n$ starts dropping towards zero, 
$a\propto t^{2/3}$, and then  
$\Omega_\Lambda$ starts to increase. Eventually $n$ is of order $10^{-5}$  
and $\Omega_\Lambda$ is of 
order 1. This type of behaviour occurs for a large, non finely tuned, 
 region of couplings $\omega$ and initial conditions. } 
\label{fig1b} 
\end{figure}

Now, why is the change in $c$ of the right order of magnitude to explain the 
results of \cite{webb}? With a solution of the form $c=c_0a^{n(t)}$ we 
find that  
\begin{equation} 
n(t)\approx {\frac{\omega \rho _\gamma }{3(\rho _m+2\rho _\Lambda )}} 
\end{equation} 
With $\omega \approx -4$ we therefore have  
\begin{equation} 
n(t_0)\approx -{\frac 43}{\frac{2.3\times 10^{-5}}{h^2(1+\Omega _\Lambda )}} 
\end{equation} 
of the right order of magnitude. The order of magnitude of the index $n\sim 
10^{-5}$, observed by \cite{webb}, is therefore fixed by the ratio of the 
radiation and the matter energy densities today. 

In summary in this theory 
gravitational effect of the 
pressure drives changes in $c$, and these convert the energy density in $%
\Lambda $ into radiation. Thus $\Lambda $ is prevented from dominating the 
universe during the radiation epoch. As the universe cools down, massive 
particles eventually become the source of pressureless matter and create a 
matter-dominated epoch. In the matter epoch the variation in $c$ comes to a 
halt, with residual effects at $z\approx 1-5$ at the level observed by Webb 
et al. As the $c$ variation is switched off, the $\Lambda $ stress 
resurfaces, and dominates the universe for a few expansion times in the 
matter-dominated era, in agreement with the supernovae results.

\section{A bit of wishful thinking}
From all that was said above it should be clear that these ideas require
and merit further work. Here we summarise what one may expect in the near future.

Obtaining a nonsingular cosmological model remains one of the outstanding
goals of string cosmology. However, it will be crucial to go beyond the
effective field theory models discussed in Section 2, and of the perturbative 
string framework used in Section 3. The recent developments in string/M theory may shed a new light on the singularity issue. The new fundamental degrees of
freedom in string theory associated with d-branes will inevitably play a
crucial role in cosmology; for the singularity problem, the dynamics of expansion and compactification, and for many other issues. We should expect
major progress in this area in the coming years.

While VSL scenarios remain an interesting idea in cosmology, it is
clear that a lot of work on its foundations needs to be done. 
The model requires breaking Lorentz symmetry and
general covariance. Clearly this is far from desirable. Moffat's early 
formulation \cite{Moffat}, relies on spontaneous, rather than explicit,
breaking of Lorentz symmetry. The idea is that the vacuum itself is not
Lorentz invariant, even though the full theory is. More recently Moffat 
and Clayton \cite{cly1,cly2} and Drummond \cite{drummond} proposed 
a bimetric formulation  for VSL. In these models the light-cone
of gravity and matter are different. These models are on firmer
foundations than the so-called Albrecht-Magueijo model. Unfortunately
their application to cosmology is rather cumbersome.  In some recent
work, it has been suggested that VSL might be realized in the brane-world.
If the Universe is a 4 dimensional brane in the process of being absorbed
by a black hole, perhaps the projected theory of gravity might resemble
the Albrecht and Magueijo theory \cite{kirt,stephon}. 
Notice that in these theories one does
not need to postulate a dynamics for the field $\psi=\log (c/c_0)$. The
dynamics of $c$ follows from the kinematics of the brane moving in the space
of the black hole. 

An  orthogonal approach has been pursued by Manida \cite{man}
and Stepanov \cite{step}. These authors have noted that breaking
Lorentz invariance does not imply breaking the principle of relativity;
merely that the constancy of the speed of light must be abandoned.
The symmetry transformation replacing Lorentz symmetry in such circumstances
is the so-called projective Lorentz transformation (or Fock-Lorentz
transformation). The ensuing theory allows for a varying speed of light
without favouring one inertial frame over another. So far this approach
has not incorporated gravity; a necessary ingredient before a realistic
application to cosmology may be discussed.

The purpose of \cite{AM98} was to show how far reaching VSL can be.
The challenge is now to set up a fundamental theory is which the model
might be realized. To our mind, the challenge is still open.


\begin{thebibliography}{10}  


\bibitem{RB99} R. Brandenberger, Inflationary Cosmology: Progress and Problems, {\it these proceedings}, hep-ph/9910410.
\bibitem{AFG} F. Adams, K. Freese and A. Guth, {\it Phys. Rev.} {\bf D43}, 965 (1991).
\bibitem{BV93} A. Borde and A. Vilenkin, {\it Phys. Rev. Lett.} {\bf 72},  3305  (1993).
\bibitem{BV89} R. Brandenberger and C. Vafa, {\it Nucl. Phys.} {\bf B316},  391  (1989).
\bibitem{AM98} A. Albrecht and J. Magueijo, {\it Phys. Rev.} {\bf D59}, 43516 (1998). 
\bibitem{Moffat} J. Moffat, {\it Int. J. Mod. Phys.} {\bf D2}, 351 (1993).
\bibitem{Penrose1965a} R. Penrose, {\it Phys. Rev. Lett.} {\bf 14},  57  (1965).
\bibitem{Hawking1967a} S.~W. Hawking, {\it Proc. Roy. Soc. Lond.} {\bf 300},  182  (1967).
\bibitem{Linde1983b} A.~D. Linde, {\it Phys. Lett.} {\bf B129},  177  (1983).
\bibitem{GasperiniET1992b} M. Gasperini and G. Veneziano, {\it Astropart. Phys.} {\bf 1},  317  (1993).
\bibitem{Veneziano1998a} G. Veneziano, hep-th/9802057  (1998).
\bibitem{BruVen} R. Brustein and G. Veneziano, {\it Phys. Lett.} {\bf B329}, 429
(1994).
\bibitem{EastherET1995a} R. Easther, K. Maeda, and D. Wands, {\it Phys. Rev.} {\bf D53},  4247  (1996).
\bibitem{KaloperET1995a} N. Kaloper, R. Madden, and K.~A. Olive, {\it Nucl. Phys.} {\bf B452},  677  (1995).
\bibitem{KaloperET1995b} N. Kaloper, R. Madden, and K.~A. Olive, {\it Phys. Lett.} {\bf B371},  34  (1996).
\bibitem{Markov} M. Markov, {\it Pis'ma Zh. Eksp. Theor. Fiz.} {\bf 36}, 214 
(1982); \\
M. Markov, {\it Pis'ma Zh. Eksp. Theor. Fiz.} {\bf 46}, 342 (1987); 
\\
V. Ginsburg, V. Mukhanov and V. Frolov,  {\it Pis'ma Zh. Eksp. Theor. Fiz.} 
{\bf 94}, 3 (1988); \\
V. Frolov, M. Markov and V. Mukhanov, {\it Phys. Rev.} {\bf D41}, 383 
(1990).
\bibitem{MB92} V. Mukhanov and R. Brandenberger, {\it Phys. Rev. Lett.} {\bf 
68}, 1969 (1992).
\bibitem{BMS93} R. Brandenberger, V. Mukhanov and A. Sornborger, {\it Phys. 
Rev.} {\bf D48}, 1629 (1993).
\bibitem{TMB93} M. Trodden, V. Mukhanov and R. Brandenberger, {\it Phys. 
Lett.} {\bf B316}, 483 (1993).
\bibitem{BEM98} R. Brandenberger, R. Easther and J. Maia, {\it JHEP} {\bf 9808}:007 (1998), gr-qc/9806111.
\bibitem{EB99} D. Easson and R. Brandenberger, {\it JHEP} {\bf 9909}:003  (1999), hep-th/9905175.
\bibitem{KLB} N. Kaloper, A. Linde and R. Bousso, {\it Phys. Rev.} {\bf D59}, 043508 (1999).
\bibitem{AS80} A. Starobinsky, {\it Phys. Lett.} {\bf B91}, 99 (1980).
\bibitem{tdual}K. Kikkawa and M. Yamasaki, {\it Phys. Lett.} {\bf B149}, 
357 (1984); \\
N. Sakai and I. Senda, {\it Prog. Theor. Phys.} {\bf 75}, 692 (1986); \\
B. Sathiapalan, {\it Phys. Rev. Lett.} {\bf 58}, 1597 (1987); \\
P. Ginsparg and C. Vafa, {\it Nucl. Phys.} {\bf B289}, 414 (1987).
\bibitem{Hagedorn}R. Hagedorn, {\it Nuovo Cimento Suppl.} {\bf 3}, 147 
(1965).
\bibitem{MT87}D. Mitchell and N. Turok, {\it Nucl. Phys.} {\bf B294}, 
1138 (1987).
\bibitem{DST89}N. Deo, S. Jain and C.-I. Tan, {\it Phys. Rev.} {\bf 
D40}, 2626 (1989).
\bibitem{V91}G. Veneziano, {\it Phys. Lett.} {\bf B265}, 287 (1991).
\bibitem{TV92}A. Tseytlin and C. Vafa, {\it Nucl. Phys.} {\bf B372}, 
443 (1992).
\bibitem{VS94}A. Vilenkin and E.P.S. Shellard, {\it Cosmic Strings and 
other Topological Defects} (Cambridge Univ. Press, Cambridge, 1994).
\bibitem{HK95}M. Hindmarsh and T.W.B. Kibble, {\it Reprt. Prog. Phys.} {\bf 58}, 477 (1995).
\bibitem{RB94}R. Brandenberger, {\it Int. J. Mod. Phys.} {\bf A9}, 
2117 (1994).
\bibitem{MS96}M. Sakellariadou, {\it Nucl. Phys.} {\bf B468}, 319 (1996).
\bibitem{barrow}  J.D. Barrow, {\it Phys. Rev.} {\bf D59} 043515 (1999). 
\bibitem{vsl1}  J.D. Barrow and J. Magueijo, {\it Phys. Lett.} 
{\bf B443}, 104 (1998). 
\bibitem{vsl2}  J.D. Barrow and J. Magueijo, {\it Phys. Lett.} {\bf B447}, 246 (1999).
\bibitem{vsl3} J.D. Barrow and J. Magueijo, {\it Class. Quant. Grav.} {\bf 16},
1435 (1999).  
\bibitem{cly1} M. A. Clayton, J. W. Moffat, {\it Phys.Lett.} {\bf B460}, 263 (1999).
\bibitem{cly2} M. A. Clayton, J. W. Moffat, gr-qc/9910112. 
\bibitem{ot}J.D. Barrow and C. O'Toole, ``Spatial 
Variations of Fundamental Constants'', astro-ph 9904116.
\bibitem{drummond} I. T. Drummond,  gr-qc/9908058. 
\bibitem{webb}J.K. Webb, V.V. Flambaum, C.W. Churchill, 
M.J. Drinkwater and J.D. Barrow, {\it Phys. Rev. Lett.} {\bf 82}, 884 (1999).   
\bibitem{beck}J.D.Bekenstein, {\it Com. on Astroph.} {\bf VIII}, 89 (1979).
\bibitem{bek2}J.D.Bekenstein, {\it Phys. Rev.} {\bf D25}, 1527 (1982). 
\bibitem{avel} P. P. Avelino and  C. J. A. P. Martins,
 {\it Phys. Lett.} {\bf B459}, 468 (1999). 
\bibitem{cole} D. H. Coule, gr-qc/9811058. 
\bibitem{steve}S. Weinberg, ``Theories of the cosmological constant",
in {\it Critical dialogues in cosmology}, ed. Neil Turok, (World Scientific, Singapore, 1997), astro-ph/9610044.
\bibitem{gupi}
V. Mukhanov and G. Chibisov, {\it Zh. Eksp. Teor. Fiz} {\bf 83} (1982) 475.
S. W. Hawking, {\it Phys. Lett} {\bf B\thinspace115} (1982) 295. 
A. Starobinsky, {\it Phys. Lett.} {\bf B\thinspace117} (1982) 175.
A. Guth and S.-Y. Pi {\it Phys. Rev. Lett.} {\bf 49}, (1982) 1110.
J. Bardeen, P. Steinhardt, and M. Turner, {\it Phys. Rev. D} {\bf 28}
(1983) 679. 
\bibitem{Peebles} P.J.E. Peebles, {\it Principles of Physical Cosmology},
371-373 (Princeton Univ. Press, Princeton).
\bibitem{quint}  I. Zlatev, L. Wang, P. Steinhardt, {\it Phys. 
Rev. Lett.} {\bf 82} (1999) 896-899. 
\bibitem{sn}J. D. Barrow and  J. Magueijo, astro-ph/9907354. 
\bibitem{super}  S. Perlmutter et al, {\it Ap. J.} {\bf 483}, 565 
(1997); S. Perlmutter et al (The Supernova Cosmology Project), {\it Nature} 
{\bf 391} 51 (1998). 
\bibitem{super1}  P. Garnavich et al.  {\it Ap. J. (Lett.)} 
{\bf 493}, L53 (1998). 
\bibitem{super2}  B. Schmidt, {\it Ap. J.} {\bf 507}, 46 (1998). 
\bibitem{super3}  A. Riess et al., {\it Ap. J.} {\bf 116}, 1009 (1998). 
\bibitem{kirt} E. Kiritsis, hep-th/9906206.
\bibitem{stephon} S. Alexander, ``On the varying speed of light in a brane-induced FRW Universe", hep-th/9912037. 
\bibitem{man}  S. N. Manida, gr-qc/9905046.
\bibitem{step} S. S. Stepanov, physics/9909009, astro-ph/9909311.
\end{thebibliography}
\end{document}